\documentclass[aps,prd,twocolumn,amsmath,amssymb,nofootinbib]{revtex4-2}

\pdfoutput=1
\DeclareUnicodeCharacter{2212}{\textendash}
\usepackage{graphicx,subfigure}
\usepackage[export]{adjustbox}
\usepackage{dcolumn}
\usepackage{bm}
\usepackage{physics}
\usepackage{xcolor}
\usepackage{slashed,cancel}
\usepackage{multirow}
\usepackage{float}
\usepackage{capt-of}
\usepackage{comment}
\usepackage{mathrsfs}
\usepackage{enumitem}
\usepackage[colorlinks=true,linkcolor=blue,urlcolor=blue,citecolor=blue]{hyperref}

\usepackage{tikz}
\usepackage{lmodern}
\usetikzlibrary{
  arrows.meta,
  backgrounds,
  calc,
  decorations.pathreplacing,
  positioning,
  shapes.geometric
}

\begin{document}

\title{What the Primary Lund Plane Omits: A Model-Stitching Diagnosis of a Jet-Tagger Representation Gap}

\author{Saurabh Rai}
\email{saurabhrai25@iitk.ac.in}
\author{Sanmay Ganguly}
\email{sanmay@iitk.ac.in}
\affiliation{Department of Physics, Indian Institute of Technology Kanpur, Uttar Pradesh 208016, India}

\begin{abstract}
A jet tagger built on the Lund plane and one built on a Particle Transformer (ParT) can differ in performance for two distinct reasons. Either the representation itself is weaker, or an adequate representation is read by a weaker downstream computation. Standalone performance cannot separate these. We use model stitching to separate these directly. It links the frozen early layers of one trained network to the frozen late layers of another through a single trained linear map. As a positive control, we use a deliberately incomplete Lund representation, the primary declustering branch, whose deficit is already known independently. Stitched into ParT's own frozen downstream, the primary-branch representation reaches a background rejection of 128--136 at fixed signal efficiency. That is at least as much as its own downstream recovers from the same bottleneck (120--134). Yet it remains far below ParT's own 291. The deficit is therefore representational, not computational. A hypergraph reader attending over triplets of splittings was tested as an alternative explanation. It extracts no more from the primary branch than the pairwise-attention network does, 116 against 112. This rules out higher-order correlations as the cause. The gap is invisible in AUC, which stays within 1\% of its target value at every depth. It appears only in background rejection at low signal efficiency. It localizes to 0.6\% of background jets. These jets sit at and above the top mass, and have more constituents but fewer primary-plane splittings than typical background. Rebuilding the Lund representation on the full Cambridge/Aachen declustering tree repairs the deficit, as predicted. Standalone rejection rises from 136 to 248, closing 72\% of the gap to ParT. Masking nodes at inference shows the repair is not localized to any small subset. The full-tree network depends almost entirely on the primary branch: deleting twelve secondary nodes costs a fifth of the rejection, while deleting twelve primary nodes costs almost all of it. Yet the secondary splittings still matter. They make the representation learnable in the first place, rather than supplying information the finished network reads. Within the primary branch, importance concentrates at the widest-angle splitting. It is worth nearly twice a random one, tied to a boosted top's opening angles. The same pattern holds for quark/gluon tagging, but far more weakly. It shows that the  Casimir-scaling radiation of the primary branch alone retains 77.6\% of ParT's rejection, against 46.9\% for top tagging, and the full tree closes only 33\% of the gap, against 72\%.

\end{abstract}

\maketitle

\section{Introduction}
\label{sec:intro}

A jet tagger built on a physics-motivated representation and one built on a learned representation of raw kinematics can differ in final performance for reasons a performance comparison alone cannot distinguish. The gap could mean the physics-motivated representation carries less information than the learned one. It could also mean an adequate representation is being read by a weaker downstream computation. These are different problems with different fixes, and standalone accuracy cannot tell them apart.

The primary Lund plane \cite{Dreyer:2018nbf} re-expresses a jet as a sequence of declustering variables tied to the collinear and soft-collinear structure of perturbative QCD, following only the harder branch at each splitting. LundNet \cite{Dreyer:2020brq} instead reads the full Cambridge/Aachen declustering tree with a graph network, and outperforms taggers built on the primary branch alone. The Particle Transformer (ParT) \cite{Qu:2022mxj} keeps every constituent's four-momentum and lets self-attention \cite{attention} build whatever features the classification loss rewards.

Comparing final performance is also easy to get wrong by choice of metric. In the benchmark that introduced LundNet, its five-variable model and ParticleNet \cite{Qu:2019gqs} differ in top-tagging AUC by only 0.004 (0.987 against 0.983), while their background rejection at $\varepsilon_S = 0.5$ differs by a factor of 2.5 (5000 against 2000) \cite{Dreyer:2020brq}. AUC, the field's most-quoted metric, can be nearly blind to a difference that background rejection at a realistic operating point shows clearly. Our own measurements reproduce this insensitivity in a sharper form: the representational deficit this paper is about moves AUC by less than one percent while halving background rejection.

Representational-similarity measures such as centered kernel alignment (CKA) \cite{cka} and singular vector canonical correlation analysis (SVCCA) \cite{svcca} compare geometry directly, but geometric similarity is not functional interchangeability: two representations can occupy well-aligned subspaces and still be unusable by a specific downstream computation, or vice versa. Model stitching \cite{modelstitching} asks the functional question directly, by connecting the frozen early layers of one trained network to the frozen late layers of another through a single trained layer, and reading off how much performance survives the connection. The same freeze-and-reconnect construction already has a different use in jet tagging: freezing early layers and fine-tuning only the rest transfers a LundNet- or ParticleNet-style tagger to a new signal on an order of magnitude less data \cite{Dreyer:2022transfer}. Two concurrent studies ask what a trained jet tagger actually uses by other means: causal ablation and path patching inside ParT itself find a sparse attention-head circuit \cite{Rai:2026mech}; perturbation- and gradient-based attribution finds which Lund-plane splittings a LundNet-style tagger relies on \cite{Patel:2026explain}. Neither separates a representation's content from its reader, which is the question this paper asks instead.

Two obstacles stand in the way of applying this to ParT and a Lund-plane network. A jet's $N$ particles and its Lund-plane splittings are sequences of unrelated length, so no per-token correspondence exists. A graph network's local message passing and a transformer's global self-attention differ enough that a stitching result could reflect either the input representation or the architecture doing the reading, with no way to separate the two afterward. We remove both obstacles by construction. The stitch point is the class token each network produces, through class attention, which is a single whole-jet vector regardless of sequence length. The Lund-plane network, which we call Lund-ParT, is the same architecture as ParT in every respect except its input: the same self-attention blocks, the same class-attention readout, differing only in the per-node feature vector and the pairwise bias function. Any result is then attributable to representation content, not to a choice of architecture.

We validate this diagnostic on a case where the answer is already known. The primary Lund plane keeps only the harder branch at each splitting: a jet with 49.0 real constituents on average yields only 12.6 primary-plane splittings, discarding most of the jet's radiation. LundNet exists precisely because this branch is incomplete. We use Lund-ParT built on the primary branch as a positive control, not because we expect it to compete with the full declustering tree, but because its deficit is known independently of anything model stitching tells us. If stitching correctly identifies that deficit as representational rather than computational, localizes it to the jets it should affect, and confirms the diagnosis when the omission is repaired, that is direct evidence the method works on cases where the ground truth is not known.

The paper has six parts. Section~\ref{sec:method} sets out the architectures, the stitching protocol, and the two normalizations it supports. Section~\ref{sec:representational} shows the gap between ParT and primary-branch Lund-ParT is representational rather than downstream, using standalone performance, retained AUC, background rejection under both normalizations, and geometric alignment. Section~\ref{sec:localize} localizes the deficit to a specific, physically identifiable jet population. Section~\ref{sec:repair} rebuilds Lund-ParT on the full declustering tree and confirms the diagnosis by repairing it. Section~\ref{sec:nodes} then asks which nodes of that tree the repaired network actually uses, and finds that the repair is not carried by any small subset of them. Section~\ref{sec:endpoint} asks the same kind of question of the primary branch alone, and finds its information concentrated at the wide-angle end of the chain. Appendix~\ref{sec:appendix-qg} repeats the entire pipeline on a second task, quark/gluon tagging, to test whether the diagnosis and its conclusions are specific to top substructure.

\section{Method}
\label{sec:method}

\subsection{Dataset}

We use the Kasieczka et al.\ top-tagging reference dataset \cite{Kasieczka:2019dbj}: 1{,}211{,}000/403{,}000/404{,}000 train/validation/test jets of hadronically decaying, boosted top quarks against light-quark and gluon jets, $p_T \in [550, 650]$~GeV, up to 200 zero-padded constituent four-momenta per jet. All results here use a fixed, reproducibly seeded 50\% random subsample of each split (605{,}500/201{,}500/202{,}000 jets), of which 100{,}971 test jets are background. Standalone ParT reaches a background rejection of 291 at $\varepsilon_S = 0.5$ at this scale, still below the $\sim\!2000$--$5000$ range reported at full dataset scale for comparable architectures \cite{Dreyer:2020brq}; the residual gap is a matter of training data volume, not of the comparisons made here, since every claim in this paper is a ratio or a difference between two models trained on identical data. Jets with fewer than two real constituents have no Lund-plane splittings under either representation and are dropped identically from both before training, affecting two jets in the test split.

\subsection{Architectures}

Both networks instantiate the same \texttt{ParticleTransformer} class: an input embedding, a pairwise-bias network computed once and reused as an additive attention bias in every block, $L=6$ pre-norm self-attention blocks (embedding dimension 128, 8 heads), 2 class-attention blocks \cite{cait} in which a learned class token cross-attends to the block-6 token sequence, and a two-layer classification head. ParT's per-particle input is the seven kinematic scalars of Ref.~\cite{Qu:2022mxj}; its pairwise bias is the four relativistic invariants $(\ln k_T, \ln z, \ln \Delta R, \ln m^2)$ of every particle pair. This totals 1{,}658{,}814 trainable parameters.

Lund-ParT is the identical class, instantiated on the primary Lund plane's five declustering coordinates as both the per-node input and the pairwise-bias basis. The primary Lund plane is built by reclustering a jet's own constituents with the Cambridge/Aachen algorithm \cite{Dokshitzer:1997in,Wobisch:1998wt} at an effective radius of 1000, using FastJet \cite{Cacciari:2011ma}, then walking only the harder branch at every splitting. For splitting $i$, parent $\to$ harder subjet $a$ + softer subjet $b$,
\begin{align}
z^{(i)} &= \frac{p_T^b}{p_T^a + p_T^b}, \nonumber \\
\Delta R^{(i)} &= \sqrt{(y^a - y^b)^2 + (\phi^a - \phi^b)^2}, \nonumber \\
k_T^{(i)} &= p_T^b\, \Delta R^{(i)}, \nonumber \\
m^{(i)} &= |(a+b)|, \nonumber \\
\psi^{(i)} &= \arctan \frac{y^b - y^a}{\phi^b - \phi^a},
\end{align}
giving the node feature $x_i = (\ln z, \ln \Delta R, \psi, \ln m, \ln k_T)$. This is the primary branch only, following Ref.~\cite{Dreyer:2018nbf}'s original construction. LundNet's own graph \cite{Dreyer:2020brq} uses the full declustering tree instead, the difference Sec.~\ref{sec:repair} tests directly. Lund-ParT's pairwise bias is the raw five-dimensional coordinate difference $x_i - x_j$. Lund-ParT has 1{,}658{,}616 trainable parameters, matching ParT's count to within 0.01\%. Background jets in the test split carry 43.3 real particles on average and yield 12.6 primary-Lund-plane splittings, since following only the harder branch discards most constituents from the resulting sequence.

Both networks are trained independently for five random seeds: AdamW \cite{adamw}, cross-entropy loss, up to 30 epochs with early stopping on validation AUC (patience 5), batch size 256, float16 mixed precision with dynamic loss scaling. Standalone test performance is in Table~\ref{tab:standalone}.

\begin{table*}[t]
\caption{Standalone test performance, mean $\pm$ standard deviation over 5 seeds, 50\% data scale. The full-tree column anticipates Sec.~\ref{sec:repair}.}
\label{tab:standalone}
\begin{ruledtabular}
\begin{tabular}{lccc}
 & ParT & Lund-ParT (primary) & Lund-ParT (full tree) \\
\hline
AUC & $0.9838 \pm 0.0003$ & $0.9759 \pm 0.0004$ & $0.9822 \pm 0.0001$ \\
rej.\ at $\varepsilon_S = 0.5$ & $291 \pm 13$ & $136 \pm 4$ & $248 \pm 5$ \\
rej.\ at $\varepsilon_S = 0.3$ & $926 \pm 46$ & $495 \pm 10$ & $895 \pm 58$ \\
\end{tabular}
\end{ruledtabular}
\end{table*}

\subsection{Stitching}
\label{sec:stitchmethod}

\begin{figure*}[t]
\centering
\includegraphics[width=0.95\textwidth]{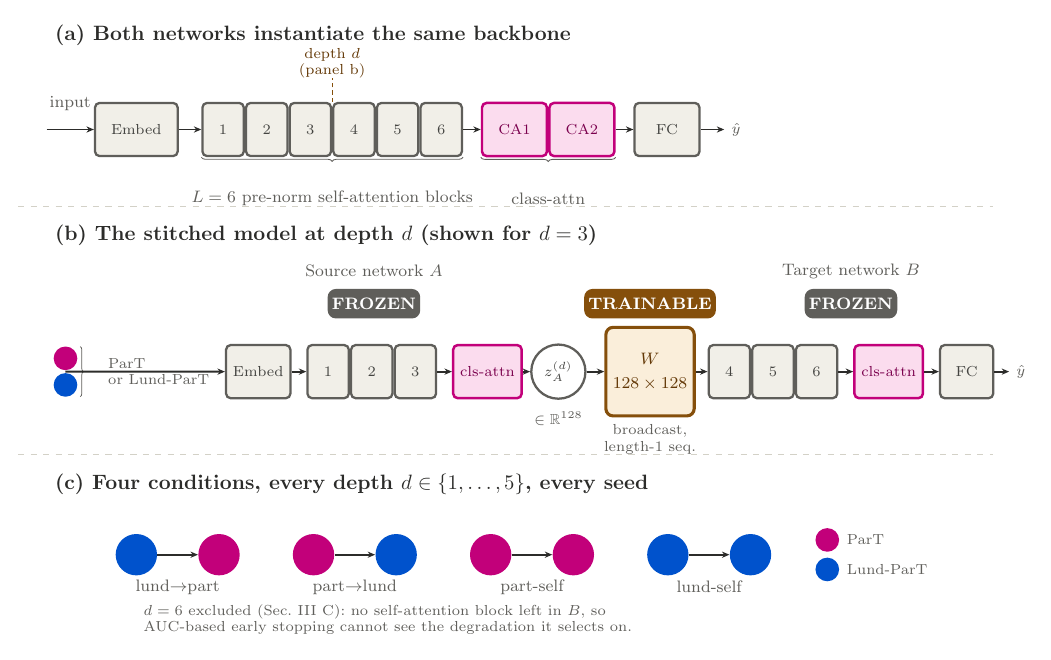}
\caption{Schematic of the stitching construction. (a) Both ParT and Lund-ParT instantiate the same six-block backbone with a two-block class-attention readout; the dashed line marks where a stitch cuts at depth $d$. (b) At depth $d$ (shown for $d=3$), source network $A$ contributes its first $d$ blocks and class attention readout, frozen at standalone trained values, producing a single class token $z_A^{(d)} \in \mathbb{R}^{128}$; a trained $128\times128$ linear map $W$ carries it, broadcast as an unit length sequence, into target network $B$'s remaining blocks, readout, and head, likewise frozen. (c) The four conditions run at every depth and seed: both cross-architecture directions and the two same-architecture self stitching controls; $d=6$ is excluded for the reason given below.}
\label{fig:stitchmechanism}
\end{figure*}

For network $N \in \{A,B\}$ and depth $d$, define
\begin{align}
f_N^{(d)} &= \mathrm{ca}_N \circ \mathrm{block}_N^{(d)} \circ \cdots \circ \mathrm{block}_N^{(1)} \circ \mathrm{embed}_N, \nonumber \\
g_N^{(d)} &= \mathrm{fc}_N \circ \mathrm{ca}_N \circ \mathrm{block}_N^{(6)} \circ \cdots \circ \mathrm{block}_N^{(d+1)},
\end{align}
so that $f_N^{(d)}: \mathcal{X} \to \mathbb{R}^{128}$ carries a jet to $N$'s class token after $d$ self-attention blocks, and $g_N^{(d)}: \mathbb{R}^{128} \to \mathbb{R}$ carries an injected 128-dimensional vector through $N$'s remaining blocks, class-attention readout, and head to a logit. For source network $A$, target network $B$, and depth $d$, the stitched classifier is then (Fig.~\ref{fig:stitchmechanism})
\begin{equation}
\hat y_{A \to B}^{(d)}(x) \;=\; g_B^{(d)}\!\big(W f_A^{(d)}(x)\big), \qquad W \in \mathbb{R}^{128 \times 128},
\end{equation}
with $f_A^{(d)}$ and $g_B^{(d)}$ fixed at their standalone-trained values and only $W$ optimized.

$A$'s embedding, first $d$ blocks, and class-attention readout are frozen at standalone-trained values and produce $A$'s own class token at depth $d$. $W$ is a single trained $\mathrm{Linear}(128,128)$ layer (16{,}512 parameters). $B$'s remaining blocks, class-attention readout, and classification head stay frozen at their own standalone trained values. Only $W$ is optimized: AdamW, up to 15 epochs with early stopping (patience 4), on a fixed random  training subsample with 20{,}000 jets. A map with 16{,}512 parameters
does not need the full training set to converge. All reported test metrics use the full 50\% test split.

$A$'s and $B$'s token sequences have unrelated lengths, so $B$'s remaining blocks run on the single injected vector $Wf_A^{(d)}(x)$ as an unit length sequence. This is exact, not approximate. Write a block's self-attention sublayer as $\mathrm{Attn}(q_i, \{k_j\}, \{v_j\}) = \sum_j \mathrm{softmax}_j(q_i \cdot k_j / \sqrt{d_k})\, v_j$. If the input sequence consists of $n$ identical tokens, $x_j = v$ for every $j = 1,\dots,n$, then every query and key is likewise identical, $q_i \equiv q$ and $k_j \equiv k$, and the softmax weights become uniform:
\begin{equation}
\mathrm{Attn}\big(q, \{k\}_{j=1}^n, \{v\}_{j=1}^n\big) \;=\; \sum_{j=1}^n \frac{e^{\,q \cdot k / \sqrt{d_k}}}{\sum_{j'=1}^n e^{\,q \cdot k / \sqrt{d_k}}}\, v \;=\; v,
\end{equation}
independent of $q$, $k$, and $n$. The point wise feedforward sublayer acts identically on every position, so a block maps $n$ identical input tokens to $n$ identical output tokens; by induction over blocks $d+1,\dots,6$, $g_B^{(d)}$ applied to $n$ copies of $Wf_A^{(d)}(x)$ returns the same value for every $n \ge 1$, including $n=1$. Broadcasting the injected vector to any number of positions therefore computes the identical function as feeding it once; we use $n=1$. Whether an unit length input costs anything relative to the token sequences $B$'s blocks were trained on is a separate question, answered empirically by the same-architecture self-stitching controls described next: self-stitched retention ranges from 29\% to 99\% of standalone performance across depths (Sec.~\ref{sec:representational}). We do not argue this cost away. Every comparison in this paper is between a cross-architecture condition and a same-architecture self-stitch at the same depth and the same source, so the injection cost is common to both sides and cancels.

We report $d \in \{1, \dots, 5\}$ and exclude $d = L = 6$ entirely. At $d=6$ the target contributes no self attention blocks at all: the construction reduces to feeding $A$'s complete, already-pooled class token through $W$ directly into $B$'s frozen class attention readout and classification head. That measures whether a linear map can re-encode a finished jet summary for a foreign classifier head, which is a different question from the one the depth sweep asks, and it introduces a double class attention pooling with no counterpart at any other depth. It also behaves pathologically in a way that makes the distinction concrete: at $d=6$ the two conditions whose target is Lund-ParT lose roughly half their background rejection while their AUC moves by less than 0.005, with a per seed scatter spanning a factor of three, whereas the two conditions whose target is ParT are unaffected. Since early stopping selects on validation AUC, and AUC is flat from the first epoch at this depth, the selection criterion cannot see the degradation it is selecting among. We therefore treat $d=6$ as outside the construction rather than as a data point to be interpreted, and it appears in no figure or table below.

Four conditions are run at every depth and every seed: $A = \mathrm{ParT}$, $B = \mathrm{Lund\text{-}ParT}$; the reverse; and the two self-stitching controls $A = B = \mathrm{ParT}$, $A = B = \mathrm{Lund\text{-}ParT}$. Every condition uses the same five seeds as standalone training, seed-matched: seed $s$'s stitching conditions use exactly seed $s$'s two standalone backbones.

\subsection{Metrics and normalization}

We report AUC and background rejection $1/\varepsilon_B$ at fixed signal efficiency $\varepsilon_S \in \{0.5, 0.3\}$. For a stitched condition with source $A$ and target $B$, we define two retained fractions for each metric $X$,
\begin{equation}
R^{\mathrm{tgt}}_X(d) = \frac{X_{\mathrm{stitched}}(d)}{X_{B,\mathrm{standalone}}}, \qquad
R^{\mathrm{src}}_X(d) = \frac{X_{\mathrm{stitched}}(d)}{X_{A,\mathrm{standalone}}}.
\end{equation}
$R^{\mathrm{tgt}}_X$ asks whether $B$'s downstream computation is the limiting factor, holding the reference $B$ fixed. $R^{\mathrm{src}}_X$ asks how much of $A$'s own information survived being read by a foreign downstream, relative to how much $A$'s own downstream recovers from the same bottleneck. These measure different things and are not interchangeable: a comparison across two conditions must hold one of $A$ or $B$ fixed and use the matching normalization, without ever mixing the standalone values of the respective targets.
Section~\ref{sec:representational} uses $R^{\mathrm{src}}_X$ throughout, since every comparison there holds the target fixed at ParT or at Lund-ParT and asks how much of a given source's information survives. We report absolute rejection alongside every fraction so the denominator is always visible.

\section{The gap is representational, not downstream}
\label{sec:representational}

Table~\ref{tab:standalone} shows the naive gap: Lund-ParT's standalone rejection, 136, is less than half of ParT's, 291. This gap is consistent with either a weaker representation or a weaker downstream computation reading an otherwise adequate one. The rest of this section separates the two.

\subsection{AUC does not distinguish the representations}

Table~\ref{tab:auc} shows retained AUC, target-normalized, for all four stitching conditions. It sits in a narrow band, 0.977--1.007, at every depth, for both cross-architecture directions and both self-stitching controls alike. The lund-to-part direction is flat to three decimal places across the entire sweep, moving from 0.991 at $d=1$ to 0.992 at $d=5$. Cross-architecture stitching is very nearly indistinguishable from same-architecture self-stitching at this metric: AUC alone would suggest the two representations are interchangeable, echoing the AUC-insensitivity noted in Sec.~\ref{sec:intro} for full taggers. Background rejection at fixed, low signal efficiency shows this picture is incomplete.

\begin{table}[t]
\caption{Retained AUC, target-normalized, primary-branch run, mean over 5 seeds.}
\label{tab:auc}
\begin{ruledtabular}
\begin{tabular}{lccccc}
 & \multicolumn{5}{c}{$d$} \\
direction & 1 & 2 & 3 & 4 & 5 \\
\hline
lund$\to$part & 0.991 & 0.991 & 0.992 & 0.992 & 0.992 \\
part$\to$lund & 0.977 & 0.996 & 1.000 & 1.004 & 1.007 \\
part-self & 0.983 & 0.989 & 0.994 & 0.996 & 0.999 \\
lund-self & 0.998 & 0.999 & 0.999 & 1.000 & 1.000 \\
\end{tabular}
\end{ruledtabular}
\end{table}

\subsection{Background rejection under both normalizations}
\label{sec:bothnorm}

\begin{figure*}[t]
\centering
\includegraphics[width=0.9\textwidth]{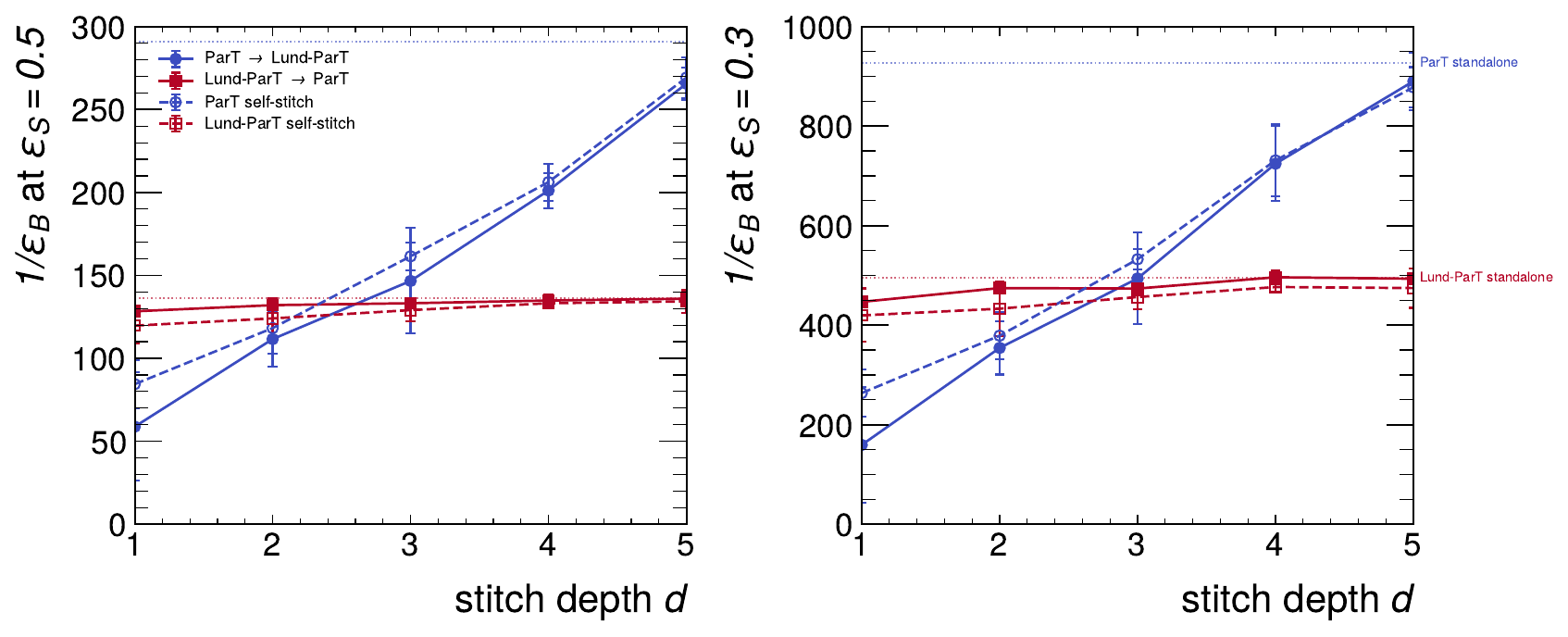}
\caption{Absolute background rejection at $\varepsilon_S = 0.5$ and $0.3$ for all four stitching conditions, primary-branch Lund-ParT, versus depth. Horizontal lines mark each network's own standalone value. The lund-to-part curve (solid red) sits at or above Lund-ParT's own self-stitching curve (dashed red) at every depth.}
\label{fig:absolute}
\end{figure*}

Figure~\ref{fig:absolute} shows absolute rejection for all four conditions at every depth, with each network's own standalone value marked. Lund-to-part -- Lund-ParT's representation read through ParT's frozen downstream -- rises gently from 128 at $d=1$ to 136 at $d=5$. Lund-self -- Lund-ParT's own downstream reading the same representation -- sits at or below it at every depth, from 120 at $d=1$ to 134 at $d=5$ (Table~\ref{tab:normalization}, both columns against the same denominator, Lund-ParT's own standalone value of 136). ParT's downstream extracts at least as much from Lund-ParT's representation as Lund-ParT's own downstream does, and slightly more at every depth. There is no latent information in the primary-branch representation that a stronger downstream would have recovered: the shortfall relative to ParT's own 291 is not explained by Lund-ParT's downstream being the bottleneck.

\begin{table}[t]
\caption{Retained rej.\ at $\varepsilon_S = 0.5$, primary-branch run, source-normalized ($R^{\mathrm{src}}$), mean over 5 seeds. lund$\to$part and lund-self both divide by Lund-ParT's own standalone value (136); part$\to$lund and part-self both divide by ParT's own standalone value (291). Self-stitching values coincide with $R^{\mathrm{tgt}}$.}
\label{tab:normalization}
\begin{ruledtabular}
\begin{tabular}{ccccc}
$d$ & lund$\to$part & lund-self & part$\to$lund & part-self \\
 & (/136) & (/136) & (/291) & (/291) \\
\hline
1 & 0.942 & 0.878 & 0.202 & 0.290 \\
2 & 0.969 & 0.910 & 0.384 & 0.407 \\
3 & 0.977 & 0.947 & 0.504 & 0.555 \\
4 & 0.990 & 0.977 & 0.692 & 0.709 \\
5 & 0.998 & 0.986 & 0.913 & 0.926 \\
\end{tabular}
\end{ruledtabular}
\end{table}

The reverse direction, part-to-lund, shows the same ordering with the roles exchanged: normalized to ParT's own standalone value (291), it retains 20--91\% across the sweep, below what ParT's own downstream recovers from the same bottleneck at every depth (29--93\%, part-self). Lund-ParT's downstream is, on this evidence, the less capable consumer of the two, independent of which representation it is given. We tested this directly rather than reading it off ten dependent per-depth comparisons: for each of the five seeds, we averaged the self-minus-cross difference in raw rejection units over $d=1$--5 separately for each direction, then averaged the two directions into one seed-level score. A one-sample $t$-test on the five resulting scores gives a mean advantage of $+8 \pm 7$ rejection units for ParT's downstream over Lund-ParT's ($t=2.56$, $p=0.062$): the same direction as every one of the ten individual comparisons, but not significant at the conventional threshold with five seeds. We report this as suggestive, not established.

This consumer-side pattern does not explain the result this section is built on. Lund-ParT's primary-branch front end is weaker than ParT's in absolute terms (136 against 291) independent of which downstream reads it, and the comparison that isolates this -- ParT's own downstream reading Lund-ParT's representation, against Lund-ParT's own downstream reading the same representation -- holds the representation fixed and varies only the reader. ParT's downstream does at least as well with it at every depth. The deficit is in what the primary branch supplies, not in who reads it.

\subsection{Geometric alignment between the two representations}

\begin{figure}[t]
\centering
\includegraphics[width=0.85\linewidth]{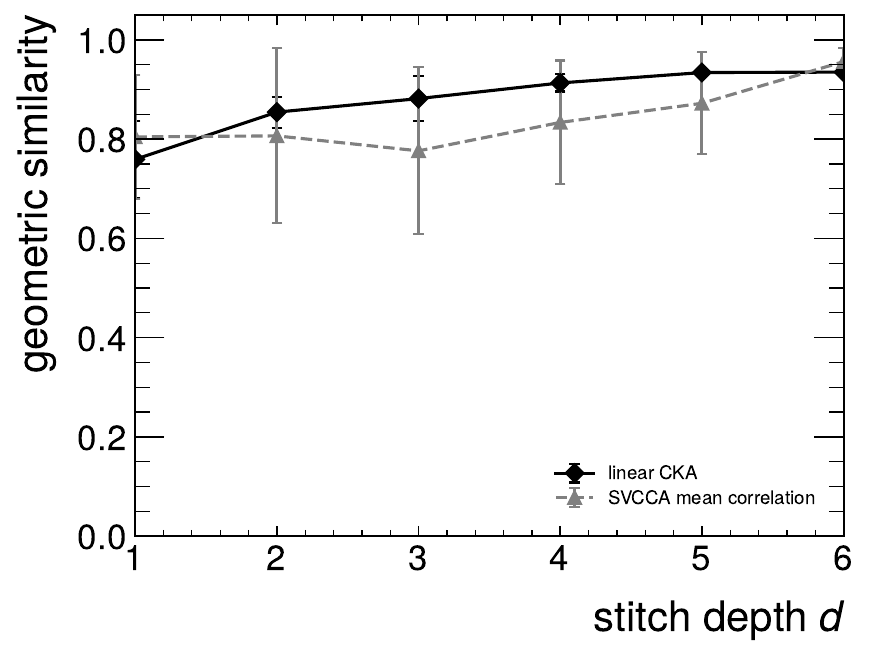}
\caption{Linear CKA and SVCCA mean canonical correlation between ParT's and Lund-ParT's class tokens versus depth.}
\label{fig:cka}
\end{figure}

Linear CKA \cite{cka} and SVCCA \cite{svcca} between ParT's and Lund-ParT's class tokens, computed independently of any stitching or classification objective on a fixed 5000-jet probe sample, both rise with depth: CKA from $0.759 \pm 0.077$ at $d=1$ to $0.934 \pm 0.005$ at $d=5$; SVCCA's mean canonical correlation from $0.804 \pm 0.126$ to $0.872 \pm 0.103$ (Fig.~\ref{fig:cka}). The depth index here labels the layer at which a class token is read from a \emph{standalone} network, so unlike the stitching sweep it remains well defined at $d=6$, where CKA reaches $0.935 \pm 0.004$; we quote the $d \le 5$ range for comparability with the rest of the paper. SVCCA's variance-retention step needed only 1--2 principal components at every depth and seed to characterize either representation, consistent with terminal-phase neural collapse in deep classifiers generally \cite{neuralcollapse}.

Geometric alignment rising with depth does not track functional transfer: lund-to-part's retained AUC (Table~\ref{tab:auc}) is flat at 0.991--0.992 across the same depth range where CKA rises from 0.76 to 0.93, and its absolute rejection gains only 8 units over that range while remaining less than half of ParT's own. The two front ends occupy increasingly similar subspaces at greater depth, and this similarity has little visible effect on how much of Lund-ParT's information ParT's downstream can extract. Geometric similarity and functional interchangeability are different properties. This is the practical case for model stitching over a representational-similarity measure alone: geometry describes the representations, stitching tests what a specific downstream computation can do with them.

\section{What the representation is missing}
\label{sec:localize}

Section~\ref{sec:representational} located the deficit in the representation. This section asks which jets are responsible, using scores averaged over five seeds for standalone ParT and the lund-to-part stitch at $d=1$ and $d=5$.

We find each model's own threshold at $\varepsilon_S = 0.5$ and compare which background jets each threshold rejects. Of the 100{,}661 background jets (of 100{,}971 total) standalone ParT rejects, the stitched model fails on $557 \pm 24$ at $d=1$ and $515 \pm 23$ at $d=5$ ($\sqrt{N}$ counting uncertainty): 0.6\% and 0.5\% of standalone ParT's rejected set. The stitched model rejects 108 background jets standalone ParT does not at $d=1$, and 107 at $d=5$. Restricted to all 100{,}971 background jets, the two models' score rankings agree closely (Spearman $\rho = 0.955$ at $d=1$, $0.956$ at $d=5$). The deficit is concentrated in a small population, not a general loss of ranking resolution.

\begin{figure*}[t]
\centering
\includegraphics[width=0.9\textwidth]{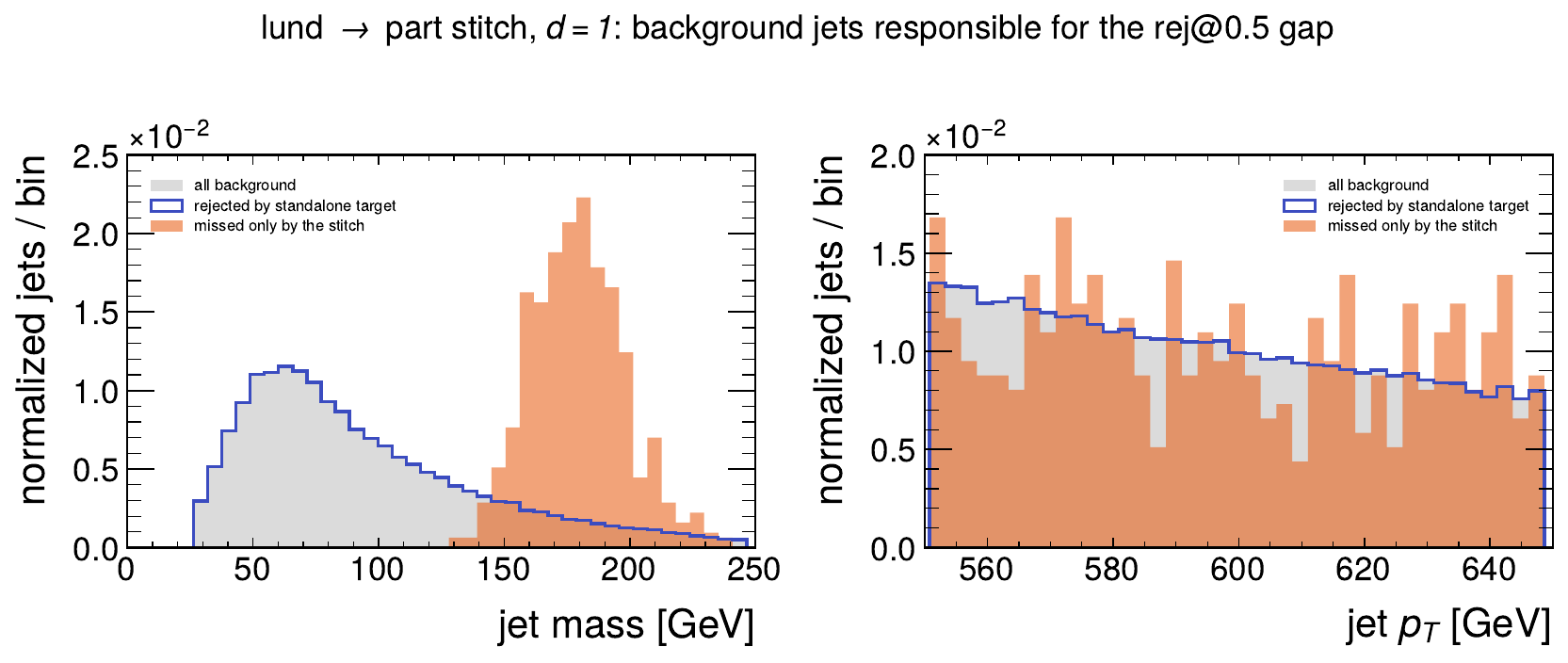}
\caption{Jet mass and $p_T$ for all background jets, the subset rejected by standalone ParT at its own $\varepsilon_S = 0.5$ threshold, and the subset the lund-to-part stitch ($d=1$) fails to reject that standalone ParT would have. The affected population concentrates at and above the top quark mass; there is no corresponding effect in $p_T$.}
\label{fig:leakage}
\end{figure*}

That population is not a random 0.6\% of standalone ParT's rejected background. Figure~\ref{fig:leakage} shows its jet-mass distribution sitting almost entirely above 150~GeV: mean 180~GeV, median 179~GeV, interquartile range 166--191~GeV, against a general background distribution with mean 95~GeV and median 82~GeV. There is no corresponding shift in $p_T$ (597~GeV against 595~GeV). These are the highest-multiplicity background jets in the sample (mean 67.2 real constituents at $d=1$ and 66.3 at $d=5$, against 43.3 typical) whose accidental invariant mass mimics a real top, distinguishable from a genuine top decay only through substructure. Despite carrying more constituents, this population has \emph{fewer} real primary-Lund-plane splittings than typical background (mean 11.3 at $d=1$ and 11.2 at $d=5$, against 12.6): more of its radiation lies off the single hardest branch the primary Lund plane records. A jet whose extra particles all hang off softer branches presents the primary plane with a shorter sequence than an ordinary jet, even though it contains more to describe. This is the diagnosis the next section tests.

\section{Confirmation: the full declustering tree}
\label{sec:repair}

If the deficit is the absence of off-primary-branch substructure, rebuilding Lund-ParT on the full C/A declustering tree, matching LundNet's own graph construction \cite{Dreyer:2020brq}, should repair it. We rebuild Lund-ParT with every splitting in the full tree as its node set, $n-1$ nodes for $n$ real constituents, mean sequence length 42.3 on background jets against the primary branch's 12.6, and repeat the identical protocol unchanged.

\begin{figure*}[t]
\centering
\includegraphics[width=0.9\textwidth]{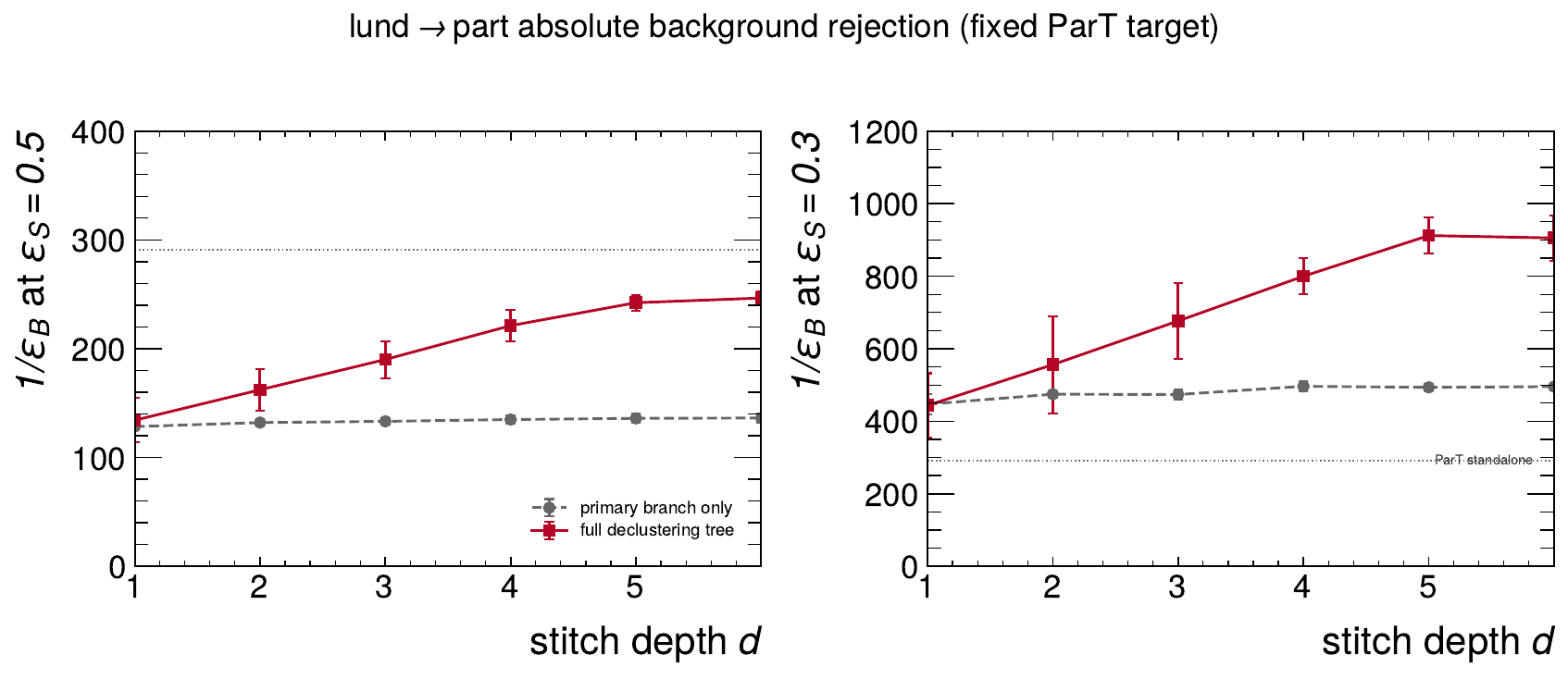}
\caption{Lund-to-part absolute background rejection at $\varepsilon_S = 0.5$ and $0.3$, primary-branch Lund-ParT versus the full-declustering-tree rebuild, same ParT target throughout. The horizontal line marks ParT's own standalone value.}
\label{fig:repair}
\end{figure*}

Standalone Lund-ParT's own rejection at $\varepsilon_S = 0.5$ rises from $136 \pm 4$ to $248 \pm 5$ (Table~\ref{tab:standalone}), closing 72\% of the distance to ParT's own $291 \pm 13$. The AUC gap closes in the same proportion but remains, as everywhere in this paper, small in absolute terms: 0.9759 to 0.9822 against ParT's 0.9838. Figure~\ref{fig:repair} compares lund-to-part's absolute rejection under both representations, against the same fixed ParT target throughout. The primary-branch curve is nearly flat, 128 to 136 across depths 1--5; the full-tree curve rises steeply from 134 at $d=1$ to 242 at $d=5$, approaching ParT's own 291. The two curves are statistically indistinguishable at $d=1$, where only one of the target's blocks has been replaced, and separate progressively as more of the Lund representation's own processing is included -- which is what a representational rather than computational difference should do.

The leaked population shrinks correspondingly, from $557 \pm 24$ jets to $399 \pm 20$ (0.4\% of standalone ParT's rejected set) at $d=1$, and from $515 \pm 23$ to $177 \pm 13$ (0.2\%) at $d=5$. The shrinkage is far larger at $d=5$, a factor of 2.9, than at $d=1$, where it is 28\%; this mirrors the separation of the two curves in Fig.~\ref{fig:repair} and has the same explanation, that at $d=1$ almost all of the computation is ParT's own and there is correspondingly little of the Lund representation in the result to repair. The Spearman rank correlation with standalone ParT rises from 0.955 to 0.975 at $d=5$, so the repair sharpens agreement in the bulk as well as in the tail.

What remains is still concentrated near the top quark mass (mean 180~GeV at $d=1$, 177~GeV at $d=5$), but its relationship to splitting count changes sign. Under the primary branch, the leaked population had fewer splittings than typical background despite more constituents; under the full tree, splitting count tracks constituent count exactly, as it must once every splitting is recorded (54.2 splittings for 55.2 particles at $d=1$; 54.0 for 55.0 at $d=5$). The signature that identified the deficit is gone, in the same jets and in the same direction the diagnosis required. Linear CKA between the two front ends also rises further under the full tree, $0.747 \pm 0.078$ at $d=1$ to $0.960 \pm 0.002$ at $d=5$, against $0.759 \to 0.934$ under the primary branch. The residual leakage in this hard-mass region is consistent with ordinary classifier disagreement, not a structurally missing quantity.

A known omission, a detected deficit localized to a specific jet population, and a repair that removes most of both: this is the validation the diagnostic needs.

\section{Which nodes carry the repair}
\label{sec:nodes}

Section~\ref{sec:repair} establishes that restoring the full declustering tree repairs the deficit. It does not say which of the roughly 36 extra nodes per jet the network uses to do it. The distinction matters practically: ``record the whole tree'' and ``record the handful of splittings that carry the discrimination'' are different prescriptions for anyone designing a Lund-based tagger, and only the second would suggest a cheaper representation is available. This section answers the question directly, by masking nodes at inference on the already-trained full-tree weights. Nothing here involves retraining.

Labelling a node as primary or secondary costs nothing, because the full-tree construction already orders its output that way. The traversal in Sec.~\ref{sec:repair} is a pre-order walk that always descends the harder subjet first, so the primary branch is exactly the leading contiguous block of a jet's node array, and its length is the node count the primary-branch representation stores for the same jet. We checked this rather than assuming it: on 2000 test jets drawn at random the leading block of the full-tree features is bit-identical to the primary-branch features, with no exceptions, and the check is repeated on every run of the analysis. Averaged over the 40{,}000 test jets used here, a jet carries $12.0$ primary and $35.9$ secondary nodes.

\begin{figure*}[t]
\centering
\includegraphics[width=0.86\textwidth]{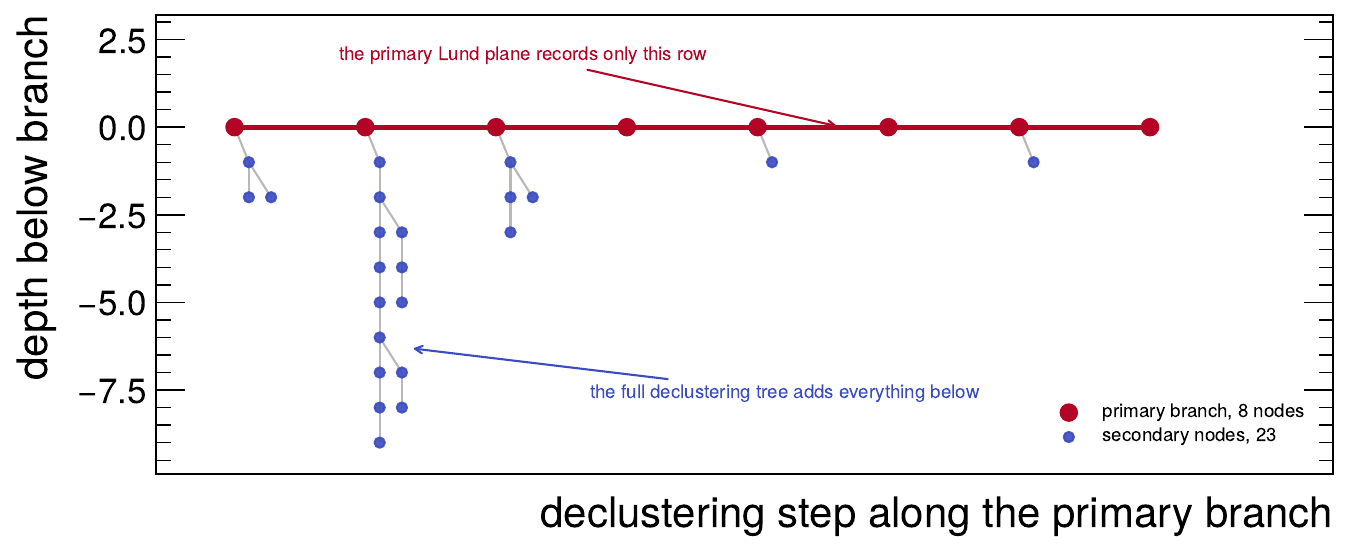}
\caption{The Cambridge/Aachen declustering tree of one real top jet. The primary branch is the horizontal spine; the subtrees hanging below it are what the primary Lund plane discards and the full tree keeps. The jet was chosen to be small enough to draw and to have a primary fraction ($0.26$) close to the population's $12.0/48.0$, but is otherwise unremarkable.}
\label{fig:treeroles}
\end{figure*}

Figure~\ref{fig:treeroles} shows what the two labels mean on a single jet. The primary branch is the spine running left to right, one node per declustering step, and every secondary node hangs somewhere below it in a subtree the primary Lund plane never records.

\begin{figure*}[t]
\centering
\includegraphics[width=0.94\textwidth]{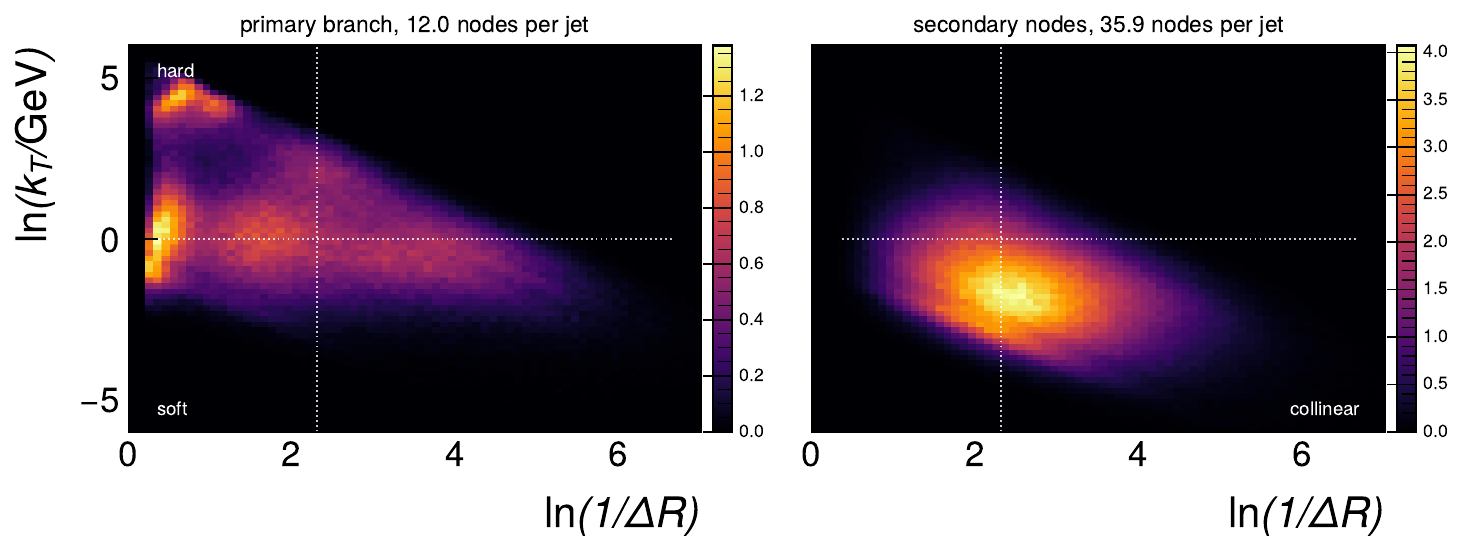}
\caption{Lund-plane density $\rho = N_{\mathrm{jet}}^{-1}\,\mathrm{d}n/\mathrm{d}\ln(1/\Delta R)\,\mathrm{d}\ln k_T$ of the two node populations, over all 40{,}000 test jets. Dotted lines mark $k_T = 1$~GeV and $\Delta R = 0.1$. The colour scales are independent, since the secondary population is three times as numerous and a shared scale leaves the left panel unreadable; the per-jet node counts in the titles fix the relative normalisation.}
\label{fig:lunddensity}
\end{figure*}

The two populations are also distinct in physics terms, as Fig.~\ref{fig:lunddensity} makes concrete. Following the harder subjet at every step is a good way to find hard structure, and the primary branch duly holds most of it: 51\% of primary nodes sit in the hard region $k_T > 1$~GeV, against 13\% of secondary nodes. Its density occupies the familiar primary-plane triangle, with a pronounced hard, wide-angle lobe. The secondary population instead concentrates in a single soft, collinear blob, 59\% of it in that one quadrant. A top decay is distinguished from a QCD jet largely by hard, wide-angle splittings, so this alone suggests the primary branch does more work than its node count implies.

The probe is a masking ablation. For each condition we suppress a chosen set of nodes and re-evaluate the same frozen weights; because the network builds its key-padding mask from the input mask and zeroes the embeddings of masked nodes, handing it a modified mask is exactly the intervention, with no change to the model. Two details matter for the result to mean anything. First, each deletion is paired with a size-matched control, so that the number of surviving tokens is held fixed and only the identity of the survivors varies; without this, any drop would conflate the loss of particular nodes with the loss of sequence length. Second, we compact the surviving nodes to the front of the sequence before the forward pass. The network trims each batch to its longest real sequence by taking a fixed number of leading columns, which is lossless for a contiguous block but silently discards scattered survivors, and a jet that loses all of them returns an undefined score. Compaction removes that asymmetry between conditions and is invisible to the network, which has no positional encoding. Jets whose every node is primary, 69 of 40{,}000, are excluded from all conditions alike, since ``delete the primary branch'' is not defined for them.

\begin{table}[t]
\caption{Background rejection and AUC under each masking condition, full-tree Lund-ParT, mean $\pm$ standard deviation over 5 seeds, 40{,}000 test jets. Conditions are grouped into matched pairs: the same number of surviving nodes, differing only in which nodes those are.}
\label{tab:ablation}
\begin{ruledtabular}
\begin{tabular}{lccc}
condition & nodes & rej.\ at $\varepsilon_S=0.5$ & AUC \\
\hline
all nodes & 48.0 & $230.6 \pm 8.2$ & $0.9818$ \\
\hline
drop 12 secondary & 36.2 & $186.5 \pm 16.3$ & $0.9789$ \\
drop 12 random & 35.9 & $53.9 \pm 3.5$ & $0.9292$ \\
drop 12 primary & 35.9 & $9.2 \pm 5.6$ & $0.8307$ \\
\hline
keep 12 primary & 12.0 & $34.7 \pm 12.4$ & $0.9229$ \\
keep 12 random & 12.0 & $4.4 \pm 1.8$ & $0.6845$ \\
\end{tabular}
\end{ruledtabular}
\end{table}

Table~\ref{tab:ablation} collects the result, and both matched pairs point the same way. Removing the twelve primary nodes costs almost everything, $230.6 \to 9.2$, while removing twelve secondary nodes costs little, $230.6 \to 186.5$, with AUC barely moving at all. Keeping only the twelve primary nodes retains more than keeping twelve random ones, $34.7$ against $4.4$. The intermediate condition is the one that shows the effect is specific rather than an artifact of deletion: twelve nodes drawn at random from the whole jet, about a quarter of which are primary by construction, land at $53.9$, between the two extremes and much closer to the primary-deletion end than their composition alone would suggest.

The comparison that carries the most weight is the first pair, because it holds the number of deleted nodes fixed at twelve and varies only their provenance: $9.2$ against $186.5$, a factor of twenty. It also disposes of the obvious objection to any ablation of this kind. Weights trained only on complete trees have never seen a shortened sequence, so a drop might reflect nothing more than that unfamiliarity. If that were the explanation, deleting twelve secondary nodes would hurt as much as deleting twelve primary ones. It does not, by a factor of twenty, so the sensitivity is to which nodes are removed and not to the fact of removal.

\begin{figure*}[t]
\centering
\includegraphics[width=0.9\textwidth]{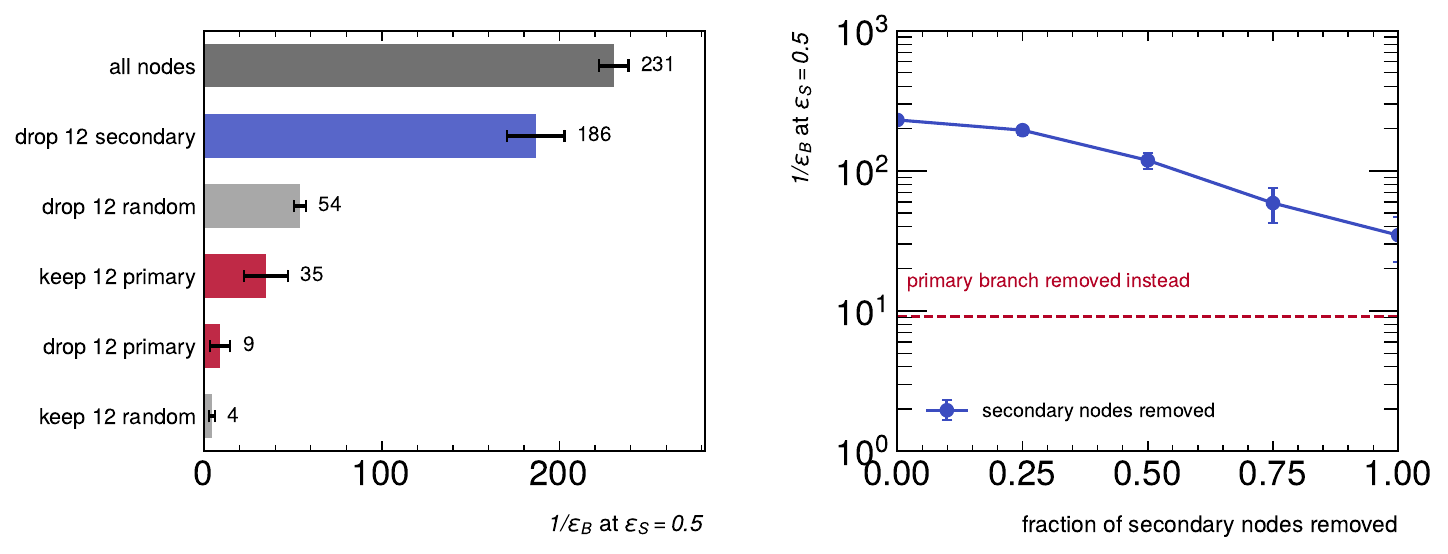}
\caption{Left: background rejection under each masking condition, with the matched pairs adjacent. Right: rejection as a growing fraction of each jet's secondary nodes is removed, ending at the primary branch alone. The dashed line marks the complementary condition, the primary branch removed and every secondary node kept.}
\label{fig:ablation}
\end{figure*}

Removing secondary nodes a fraction at a time, rather than twelve at a time, shows how their contribution is spread (Fig.~\ref{fig:ablation}, right). Rejection falls from $230.6 \pm 8.2$ with every node present to $194.9 \pm 14.2$ at a quarter removed, $118.8 \pm 15.1$ at a half, $58.9 \pm 16.1$ at three quarters, and $34.7 \pm 12.4$ when none is left. The first quarter is nearly free, costing 15\%, and each subsequent quarter costs more than the one before it. There is no step, no fraction at which the curve falls off a cliff, which is what a small load-bearing subset would have produced. The endpoint is the sharpest statement available here: stripping all 36 secondary nodes still leaves the network at $34.7$, well above the $9.2$ it reaches when the 12 primary nodes go instead and all 36 secondary ones remain.

Taken together these measurements say something more specific than Sec.~\ref{sec:repair} alone could. At inference the trained full-tree network leans overwhelmingly on the primary branch: twelve primary nodes support a higher rejection on their own, $34.7$, than all thirty-six secondary nodes do without them, $9.2$. Yet training on the full tree is what lifts standalone rejection from $136$ to $248$ in the first place. Both statements are measurements, and the way to hold them together is that the secondary nodes act during training, shaping the representation the network learns to build, rather than supplying discriminating information to be read off at test time. On that reading the repair is real but not localized, and the honest answer to which of the extra nodes carry it is that no small subset does.

This has a consequence for anyone hoping to shortcut the full tree. A representation that recorded the primary branch plus a modest selection of secondary splittings would not, on this evidence, recover the repair, because the secondary contribution is not concentrated anywhere for such a selection to capture. We would also caution against reading the individual conditions in Table~\ref{tab:ablation} as importance scores in any stronger sense. They are comparisons within matched pairs, and it is only the contrast within a pair that is interpretable; the absolute values still carry whatever cost the network pays for a sequence shorter than any it was trained on.

\section{Where along the primary branch information sits}
\label{sec:endpoint}

Section~\ref{sec:nodes} asks which nodes of the full declustering tree a trained network relies on, and finds it relies almost entirely on the primary branch. This section asks the complementary question of the primary branch itself: a trained, frozen Lund-ParT built on the primary branch alone reads a jet as a sequence of declustering steps, and we ask which single step it can least afford to lose.

Cambridge/Aachen declustering is angular ordered, so position along the chain is a proxy for angular scale. We verified this rather than assuming it: the last node sits at smaller $\Delta R$ than the first in $100.0\%$ of jets, in both top tagging and quark/gluon tagging, the latter as described in Appendix~\ref{sec:appendix-qg}, and the median $\Delta R$ falls monotonically along the chain, from $0.69$ to $0.008$ for top tagging and $0.35$ to $0.004$ for quark/gluon. The first step is therefore the widest-angle splitting and the last the most collinear.

Three ablations are compared on the already-trained, frozen primary-branch Lund-ParT, with no retraining: remove the \textbf{first} step, remove the \textbf{last}, or remove one chosen uniformly at \textbf{random}. Anchoring at the ends is what makes the comparison clean. Jets have different chain lengths ($12.0$ steps on average for top tagging, $13.6$ for quark/gluon), so ablating a fixed index would act on a shrinking and increasingly atypical subset as the index grows; every jet has a first and a last step whatever its length, so all three conditions act on $100\%$ of the jets with nothing padded, normalized or truncated. Jets with a single node cannot give one up and are excluded from every condition alike ($5$ of $201{,}998$ for top tagging, $9$ of $199{,}994$ for quark/gluon).

The random condition is the control, and it carries the argument. Removing any node costs something, partly because a shortened sequence is unfamiliar to weights trained on complete ones, so an absolute drop cannot distinguish ``this step mattered'' from ``removal hurts''. Only the \emph{paired} difference against random, computed seed by seed so that common cost cancels, is interpretable; all claims below are stated that way. Survivors are compacted to the front of the sequence before the forward pass, following Sec.~\ref{sec:nodes}'s procedure, since the network has no positional encoding and the batch trimmer would otherwise silently discard a non-contiguous mask. We report background rejection at $\varepsilon_S = 0.5$ and the threshold-free mean per-jet $|\Delta\,\text{score}|$, on cached checkpoints over five seeds.

\begin{table*}[t]
\caption{One node removed per jet from the standalone primary-branch Lund-ParT, mean $\pm$ standard deviation over 5 seeds. ``Excess'' is the paired difference against the random control, as a percentage of the unablated rejection, with its paired $t$ statistic.}
\label{tab:endpoint}
\begin{ruledtabular}
\begin{tabular}{llccc}
dataset & condition & rej.\ at $\varepsilon_S{=}0.5$ & loss & excess vs.\ random \\
\hline
top tagging & unablated & $136.28 \pm 4.26$ & --- & --- \\
 & remove first & $40.13 \pm 0.83$ & $-70.6\%$ & $+35.2\%$ ($t{=}55.8$) \\
 & remove random & $88.07 \pm 2.00$ & $-35.4\%$ & --- \\
 & remove last & $128.12 \pm 1.93$ & $-6.0\%$ & $-29.4\%$ ($t{=}-24.0$) \\
\hline
quark/gluon & unablated & $30.62 \pm 0.17$ & --- & --- \\
 & remove first & $22.32 \pm 0.26$ & $-27.1\%$ & $+11.5\%$ ($t{=}21.3$) \\
 & remove random & $25.85 \pm 0.32$ & $-15.6\%$ & --- \\
 & remove last & $29.46 \pm 0.16$ & $-3.8\%$ & $-11.8\%$ ($t{=}-20.6$) \\
\end{tabular}
\end{ruledtabular}
\end{table*}

\begin{figure*}[t]
\centering
\includegraphics[width=0.9\textwidth]{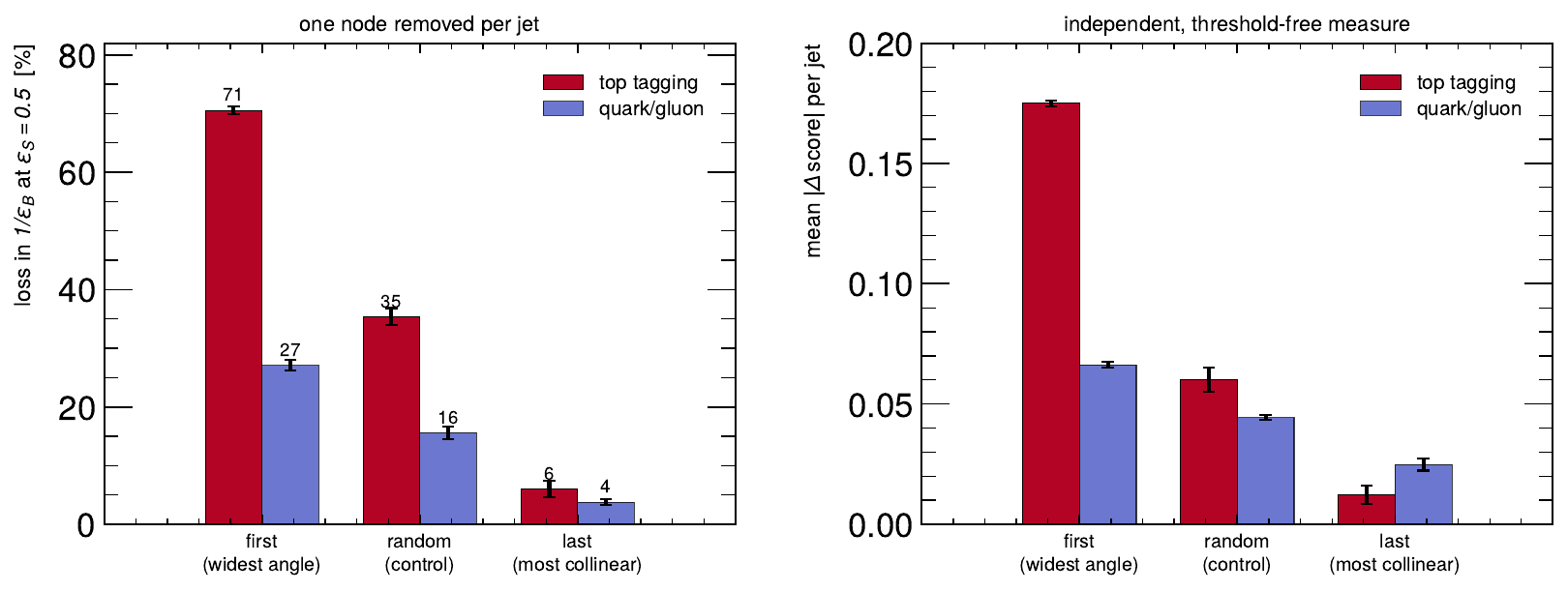}
\caption{Left: loss in background rejection when one node is removed. Right: the same conditions under the threshold-free per-jet score shift. Error bars are the seed standard deviation.}
\label{fig:endpointbars}
\end{figure*}

The widest-angle splitting is worth far more than an average one in both tasks, and about three times more so for top tagging: $+35.2\%$ of the unablated rejection against $+11.5\%$. The most collinear splitting is worth correspondingly less than average -- for quark/gluon by almost exactly the same magnitude the first splitting is worth more ($-11.8\%$ against $+11.5\%$), for top tagging by a smaller margin than the first splitting's excess ($-29.4\%$ against $+35.2\%$). Importance falls monotonically as the splitting becomes more collinear. The threshold-free measure agrees independently (Fig.~\ref{fig:endpointbars}, right): mean $|\Delta\,\text{score}|$ runs $0.175 \rightarrow 0.060 \rightarrow 0.012$ for top tagging across first, random and last, a factor of $14.6$ between the two ends, and $0.066 \rightarrow 0.044 \rightarrow 0.025$ for quark/gluon, a factor of $2.6$ -- the same ordering, the same contrast between the two tasks, with no threshold involved.

\begin{figure}[t]
\centering
\includegraphics[width=0.95\linewidth]{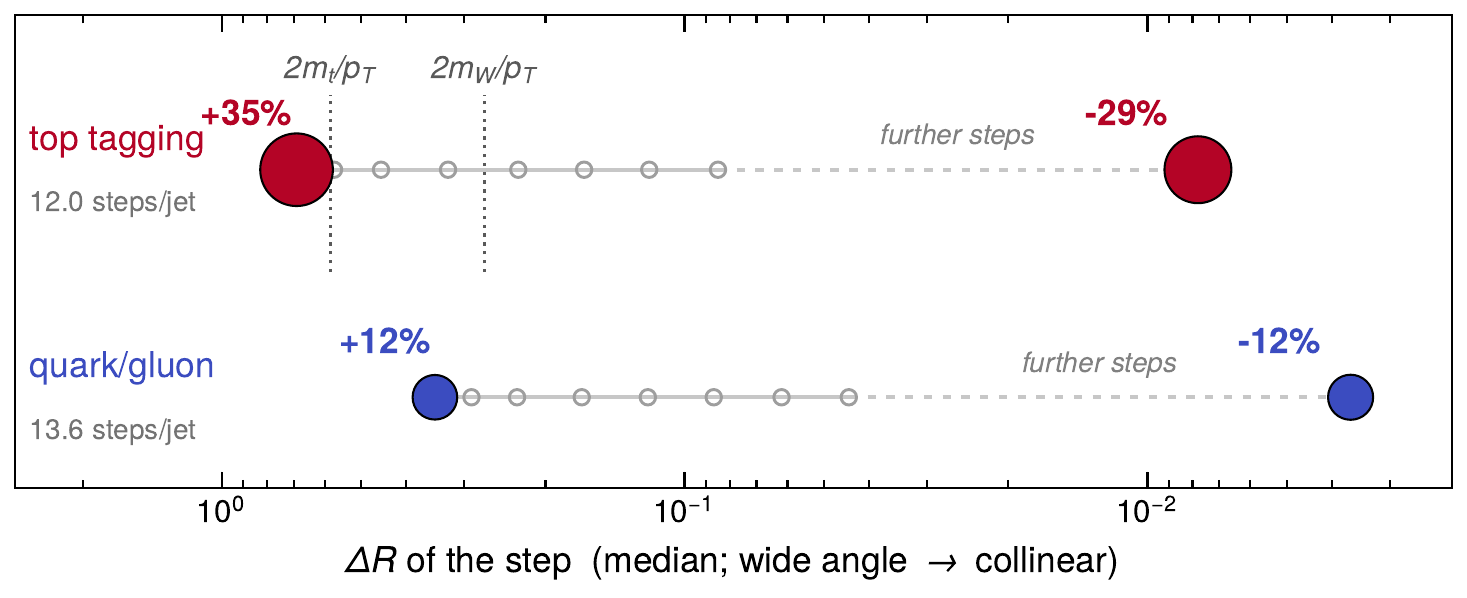}
\caption{The measured importance placed on the angular scale it corresponds to. Circles mark the median $\Delta R$ of successive declustering steps; filled circles are the two steps actually ablated, with area proportional to their excess damage over the random control, and hollow circles are the intermediate steps, which were not measured. Dotted lines mark the top and $W$ decay opening angles $2m/p_T$ at $p_T = 600$~GeV, drawn on the top-tagging lane only since they are a statement about top jets. The first splitting of a top jet falls close to these scales; quark/gluon jets have no such scale, and their importance profile is correspondingly flatter.}
\label{fig:endpointschematic}
\end{figure}

Jet-tagging information along the primary chain is concentrated at wide angles, and markedly more so for top tagging than for quark/gluon discrimination. Removing the single widest-angle splitting costs top tagging $71\%$ of its background rejection, essentially twice what removing a random splitting costs, while the most collinear splitting is nearly free to discard; the same ordering holds for quark/gluon, with roughly a third of the contrast. This has a natural reading: a boosted top is identified by a hard, localized prong structure whose opening angles $2m_t/p_T$ and $2m_W/p_T$ sit near the first declustering, so one wide-angle splitting carries much of the discrimination, while quark/gluon separation rests on Casimir scaling, which acts across the whole emission spectrum and accumulates over many steps, so no single step is decisive. This corroborates, from an independent direction, why discarding off-primary branches was close to fatal for top tagging and only mildly costly for quark/gluon (Sec.~\ref{sec:repair}, Appendix~\ref{sec:appendix-qg}).

Two limits are worth stating. This measures what a \emph{trained} network uses, not what is physically available; Sec.~\ref{sec:nodes} found the two can differ sharply. And removing any single node costs far more than its naive $1/N$ share of the chain would predict: a $1/12.0$ share is $8.3\%$ for top tagging, against the $35\%$ a random node actually costs; a $1/13.6$ share is $7.4\%$ for quark/gluon, against $16\%$. There is a real distribution-shift penalty, which is why only the paired comparison against the random control is quoted as a result.

\section{Discussion and conclusion}
\label{sec:discussion}

Model stitching, applied to ParT and a Lund-plane network built the same way, separates a performance gap into its two possible causes and identifies which one is responsible. The primary-branch representation is genuinely weaker than ParT's, independent of which downstream reads it; this shortfall is invisible in AUC and appears only in background rejection at fixed, low signal efficiency; it is concentrated in a specific, physically identifiable population of high-multiplicity, high-mass background jets with radiation lying off the primary branch; and it is repaired, in the population it should repair and in the direction it should move, by restoring exactly the branches the primary plane omits.

Section~\ref{sec:nodes} then pushes past the repair to ask what performs it. The answer complicates the natural reading of Sec.~\ref{sec:repair}. At inference the repaired network depends almost entirely on the primary branch, and the secondary nodes that the full tree adds are individually close to redundant: deleting twelve of them costs a fifth of the background rejection, while deleting twelve primary nodes costs almost all of it. The extra nodes are what makes the better representation learnable, not what the finished network reads. We think this is worth stating plainly because it is the kind of thing a performance comparison would never surface, and because it warns against the obvious engineering shortcut of keeping the primary branch plus a few chosen secondary splittings.

None of this required the primary branch to be a strong tagger on its own. It is not, by construction: LundNet uses the full declustering tree precisely because the primary branch is incomplete. We used it as a positive control because its deficit was known before any stitching result was computed, which is what makes correctly recovering that deficit, and localizing it to the right jets, evidence that the method works.

Section~\ref{sec:endpoint} asks a complementary question of the primary branch in isolation: its own discriminating power turns out to be concentrated at its wide-angle end, close to the top and $W$ opening angles, far more sharply for top tagging than for quark/gluon. That is the same picture Sec.~\ref{sec:nodes} finds from the branch side: a top jet's discrimination depends on localized, specific structure rather than being spread evenly across whatever is recorded.

\subsection{A hypergraph reading of the same representations}
\label{sec:hypergraph}

One alternative to the conclusion of Sec.~\ref{sec:representational} deserved a direct test rather than an argument. Every reader of the primary-branch representation considered so far processes its splittings through pairwise attention. A model that reasons natively over \emph{triplets} of splittings might do better: three-body correlations among recorded emissions could in principle constrain the radiation that was not recorded, letting a network infer something about the branches the primary plane discards without ever being shown them. Had that been the case, part of the deficit this paper attributes to the representation would have belonged to the reader after all.

We therefore rebuilt the tagger as a hypergraph network, in which a token is not a particle or a splitting but a triplet of them. Hyperedges are the union of a local neighbourhood construction and every triplet among the ten hardest nodes, the second included because purely local triplets sit within a single prong and miss the wide-angle structure a boosted top is identified by. The same architecture, the same jets, and the same training protocol are used throughout, with only the node set changed between the three variants. These runs use 10\% of the dataset and a single seed, so they are indicative and are not comparable with the measurements elsewhere in this paper; the companion figures in the preceding sections are untouched by them.

\begin{table}[t]
\caption{Hypergraph taggers against the corresponding stitching-study networks, all at 10\% of the dataset. The hypergraph column is a single seed and carries no uncertainty; the right-hand column is the mean over 5 seeds at the same scale.}
\label{tab:hypergraph}
\begin{ruledtabular}
\begin{tabular}{lcc}
representation & hypergraph & this paper's network \\
 & rej.\ at $\varepsilon_S=0.5$ & rej.\ at $\varepsilon_S=0.5$ \\
\hline
particles & 133 & $208 \pm 12$ \\
Lund, primary branch & 116 & $112 \pm 7$ \\
Lund, full tree & 181 & $201 \pm 11$ \\
\end{tabular}
\end{ruledtabular}
\end{table}

The test comes out negative, and cleanly so. On the primary branch the hypergraph reaches a rejection of 116 against Lund-ParT's $112 \pm 7$ at the same scale: the two agree inside one standard deviation, so the explicitly three-body reader extracts no more from that representation than the pairwise one does. The gap between the two Lund representations survives intact, 116 rising to 181 when the full tree is restored, a 55\% gain against the 79\% the stitching study measures for the same pair. A model built to see higher-order structure, given only the primary declusterings, does not recover what the full tree carries.

We draw attention to how little this buys the present study, because the temptation to present it as an extension should be resisted. Its entire value is eliminative: it removes one competing explanation for Sec.~\ref{sec:representational}'s result and leaves every number, figure and conclusion in this paper exactly as they were. It improves neither the diagnosis nor the repair, and it identifies no jet population that Sec.~\ref{sec:localize} did not already identify.

Nor can it be folded into the stitching framework without damage, which is the more important point. This paper's design rests on one architecture reading several representations, so that a stitched condition's shortfall is attributable to representation content. The hypergraph tagger breaks that symmetry: it carries no pairwise attention bias, since at roughly two hundred hyperedge tokens rather than fifty particles the bias tensor does not fit alongside training activations, and on particles it reaches only two thirds of ParT's rejection. Stitching it against ParT would therefore reintroduce precisely the confound between representation content and architecture that Sec.~\ref{sec:intro} sets out to remove, and any deficit so measured would be uninterpretable. We report the comparison as a consistency check on the representational claim and build nothing on it.

The result is scale-limited in a stated, not defensive, way. Both networks are trained at 50\% of the reference dataset, and standalone rejection here is below values reported at full scale for comparable architectures. Every claim in this paper rests on a large effect -- a factor of two in standalone rejection, a factor of 2.9 in the leaked population at the deeper repair depth -- and we would not expect this scale to change which population is responsible or the direction of the repair. Whether the same protocol, applied to a source-target pair whose representations are closer in strength, resolves depth-by-depth differences at this data scale is a separate question we do not attempt to answer here; it needs more data, more seeds, and is better reported on its own. The consumer-asymmetry result of Sec.~\ref{sec:bothnorm} is the one place where five seeds are visibly too few: the effect points the same way in all ten individual comparisons and in all five seed-level scores, yet reaches only $p=0.062$.

Two limitations of scope deserve naming. First, we excluded $d=6$ from the analysis on structural grounds (Sec.~\ref{sec:stitchmethod}) rather than reporting it with a caveat. We think this is the right call, since at that depth the target contributes no self-attention blocks and the quantity measured is not the one the sweep is designed to measure, but it does mean the sweep stops one layer short of the full stack and we cannot speak to what happens there. Second, a consumer-side fine-tuning check -- unfreezing the first target block downstream of the stitch point, to test whether the consumer asymmetry survives letting the receiving block adapt -- was performed at an earlier, smaller data scale and was inconclusive there, with a seed-dependent instability that we could not attribute to a learning-rate artifact. We have not repeated it at this scale, and we do not use it as evidence anywhere in this paper.

A natural question is whether this diagnosis is specific to top tagging, where the omitted structure is a known, localized pair of prongs. Appendix~\ref{sec:appendix-qg} repeats the full pipeline on quark/gluon tagging, a task whose discrimination rests on Casimir-scaling radiation patterns rather than localized substructure, and finds a different balance between representation and repair: the primary branch alone already retains $77.6\%$ of ParT's standalone rejection, against $46.9\%$ on top tagging, and restoring the full tree closes only $33\%$ of the remaining gap, against $72\%$. The consumer-asymmetry result of Sec.~\ref{sec:bothnorm} does not reproduce there either. The two tasks nonetheless agree on one point: a complete declustering tree sits at essentially the same fraction of ParT's rejection on both, about $85\%$, which we read as a cost intrinsic to the Lund encoding itself rather than to any specific omitted branch. The quark/gluon stitch measurements are noisier at shallow depth than the top-tagging ones, and Appendix~\ref{sec:appendix-qg} states that caveat alongside every number it affects.

Our contribution is not model stitching itself, which is due to Bansal, Nakkiran, and Barak \cite{modelstitching}. It is the class-token stitch point, which removes the sequence-length obstacle to applying stitching across architectures with unrelated per-token structure; the same-architecture Lund-ParT, which removes the confound between representation content and message-passing mechanism; the explicit dual normalization of Sec.~\ref{sec:method}; and the positive-control design that lets the method's output be checked against an independently known answer.

\section{Acknowledgments}
\label{sec:acknowledgment}
\noindent
S.R. is supported by the IIT-Kanpur institutional Ph.D fellowship.\\ S.G. is supported by the IIT-Kanpur faculty initiation grant (IITK /PHY /2023499), Anusandhan National Research Foundation -  Advanced Research Grant (ANRF/ARG/2025/005801/PS) and Anusandhan National Research Foundation - ANRF MISSION AI for Science and Engineering (ANRF/AISE/2025/000373/AISSAI).

\section*{Data Availability Statement}

The Kasieczka et al.\ top-tagging reference dataset used in this work is publicly available at Zenodo (DOI: \url{10.5281/zenodo.2603256}). Code implementing the architectures, stitching protocol, and analysis described here is available from the authors on reasonable request.

\section*{AI USAGE Statement}
The authors acknowledge the usage of Anthropic's Opus-5 model for cross checking the SVCCA analysis and code refactoring. The manuscript however is manually written by the authors themselves.

\appendix
\section{Generalization to quark/gluon tagging}
\label{sec:appendix-qg}

The representational deficit identified in Secs.~\ref{sec:representational}--\ref{sec:nodes} is specific to top tagging: the primary Lund plane discards the off-axis prongs a boosted top's three-body decay needs, but how costly that omission is for a classification task with different discriminating physics is not obvious a priori. We repeat the identical pipeline and hyperparameters -- the same architectures, the same stitching protocol, the same five seeds -- on the EnergyFlow quark/gluon dataset~\cite{Komiske:2018cqr}, a task whose discrimination rests on Casimir scaling of gluon radiation rather than on localized substructure. We report what was measured and state where it is, and is not, resolved.

\subsection{Setup}

\paragraph{Dataset.} EnergyFlow quark/gluon jets (Pythia generator, $b$- and $c$-initiated jets excluded), $1{,}000{,}000$ events, shuffled under a fixed seed and split $600{,}000/200{,}000/200{,}000$ into train, validation and test. EnergyFlow stores each constituent as $(p_T, y, \phi, \mathrm{pid})$; these were converted to four-momenta $(E, p_x, p_y, p_z)$ using each particle's PDG mass from its pid, not a massless approximation, and written into the same HDF5 layout the pipeline reads for top tagging, so both datasets are processed by identical code. Jets carry 43.2 constituents on average at a jet $p_T$ of about 523~GeV. Labels follow the EnergyFlow convention, quark $=1$ and gluon $=0$, so signal efficiency below is quark efficiency and the rejected set is gluon jets. The test split contains $100{,}016$ gluon jets.

\paragraph{Representations.} Three inputs to one architecture, as in Sec.~\ref{sec:method}: ParT takes the seven kinematic scalars per particle with the four relativistic invariants of every particle pair as its attention bias; Lund-ParT (primary) takes the five primary-Lund-plane declustering coordinates, 13.6 nodes per jet on average; Lund-ParT (full tree) takes every splitting of the full Cambridge/Aachen tree, 42.3 nodes per jet.

\paragraph{Architecture and training.} All three instantiate the same \texttt{ParticleTransformer} class used throughout this paper: 6 pre-norm self-attention blocks, embedding dimension 128, 8 heads, 2 class-attention blocks, and a two-layer head (1{,}658{,}814 parameters for ParT, 1{,}658{,}616 for Lund-ParT). Training uses AdamW, cross-entropy loss, batch size 256, up to 30 epochs with early stopping on validation AUC (patience 5), and float16 mixed precision. Five seeds throughout; every quoted uncertainty is the standard deviation over those five seeds.

\paragraph{Stitching.} Identical to Sec.~\ref{sec:stitchmethod}: for source $A$, target $B$, and depth $d$, $A$'s embedding, first $d$ blocks, and class-attention readout are frozen and produce $A$'s own class token; a single trained $\mathrm{Linear}(128,128)$ map (16{,}512 parameters) carries it across; $B$'s remaining blocks, readout, and head complete the forward pass frozen at their standalone values. Only the linear map is optimized: up to 15 epochs, patience 4, on a fixed 20{,}000-jet subsample. Four conditions are run at every depth and seed, seed-matched. Depths $d=1,\dots,5$ are reported; $d=6$ is excluded for the same structural reason given in Sec.~\ref{sec:stitchmethod}.

\paragraph{Execution.} The two Lund modes were run as separate jobs. The ParT backbones are byte-identical between them (verified by checksum), so the ParT column below is one set of five models shared by both halves rather than an independent replicate.

\subsection{Standalone performance}

\begin{table*}[t]
\caption{Standalone test performance on quark/gluon tagging, mean $\pm$ standard deviation over 5 seeds.}
\label{tab:qg-standalone}
\begin{ruledtabular}
\begin{tabular}{lccc}
 & ParT & Lund-ParT (primary) & Lund-ParT (full tree) \\
\hline
AUC & $0.9110 \pm 0.0011$ & $0.8905 \pm 0.0003$ & $0.9006 \pm 0.0007$ \\
rej.\ at $\varepsilon_S = 0.5$ & $39.44 \pm 0.73$ & $30.59 \pm 0.17$ & $33.54 \pm 0.52$ \\
rej.\ at $\varepsilon_S = 0.3$ & $98.14 \pm 2.52$ & $82.13 \pm 0.92$ & $88.12 \pm 1.55$ \\
\end{tabular}
\end{ruledtabular}
\end{table*}

Restoring the full declustering tree moves standalone rejection from $30.59$ to $33.54$ against ParT's $39.44$ (Table~\ref{tab:qg-standalone}), closing $33.3\%$ of the gap -- far less than the $72\%$ closed on top tagging (Table~\ref{tab:standalone}).

\subsection{Stitched performance}

Table~\ref{tab:qg-absolute} gives the absolute rejection of the lund-to-part stitch, the quantity through which the two Lund representations are compared, with the same frozen ParT target in both rows.

\begin{table*}[t]
\caption{Absolute background rejection at $\varepsilon_S = 0.5$ for the lund-to-part stitch on quark/gluon tagging, both Lund representations, against the same ParT target. ParT's own standalone value is $39.44$.}
\label{tab:qg-absolute}
\begin{ruledtabular}
\begin{tabular}{lccccc}
 & \multicolumn{5}{c}{$d$} \\
representation & 1 & 2 & 3 & 4 & 5 \\
\hline
primary branch & $15.1 \pm 8.2$ & $21.9 \pm 7.0$ & $25.0 \pm 4.6$ & $23.6 \pm 12.4$ & $30.1 \pm 0.3$ \\
full tree & $7.8 \pm 7.1$ & $15.2 \pm 7.9$ & $20.8 \pm 6.0$ & $30.6 \pm 0.8$ & $32.8 \pm 0.7$ \\
\end{tabular}
\end{ruledtabular}
\end{table*}

\begin{table*}[t]
\caption{Retained AUC, target-normalized ($R^{\mathrm{tgt}}$), primary-branch run, quark/gluon tagging.}
\label{tab:qg-auc}
\begin{ruledtabular}
\begin{tabular}{lccccc}
 & \multicolumn{5}{c}{$d$} \\
direction & 1 & 2 & 3 & 4 & 5 \\
\hline
lund$\to$part & $0.930 \pm 0.053$ & $0.963 \pm 0.016$ & $0.969 \pm 0.008$ & $0.858 \pm 0.263$ & $0.977 \pm 0.002$ \\
part$\to$lund & $0.941 \pm 0.066$ & $0.839 \pm 0.341$ & $0.989 \pm 0.022$ & $1.004 \pm 0.007$ & $1.016 \pm 0.006$ \\
part-self & $0.931 \pm 0.045$ & $0.953 \pm 0.035$ & $0.964 \pm 0.021$ & $0.981 \pm 0.008$ & $0.993 \pm 0.004$ \\
lund-self & $0.970 \pm 0.026$ & $0.981 \pm 0.018$ & $0.994 \pm 0.002$ & $0.997 \pm 0.001$ & $0.998 \pm 0.003$ \\
\end{tabular}
\end{ruledtabular}
\end{table*}

\begin{table*}[t]
\caption{Retained rejection at $\varepsilon_S = 0.5$, source-normalized ($R^{\mathrm{src}}$), primary-branch run, quark/gluon tagging. lund$\to$part and lund-self divide by Lund-ParT's standalone $30.59$; part$\to$lund and part-self divide by ParT's $39.44$. Uncertainties propagate the numerator's seed spread only; the denominators are known to $0.6\%$ and $1.9\%$ respectively.}
\label{tab:qg-rsrc}
\begin{ruledtabular}
\begin{tabular}{lccccc}
 & \multicolumn{5}{c}{$d$} \\
direction & 1 & 2 & 3 & 4 & 5 \\
\hline
lund$\to$part & $0.494 \pm 0.268$ & $0.715 \pm 0.228$ & $0.818 \pm 0.149$ & $0.772 \pm 0.405$ & $0.983 \pm 0.011$ \\
lund-self & $0.547 \pm 0.259$ & $0.655 \pm 0.234$ & $0.858 \pm 0.097$ & $0.944 \pm 0.038$ & $0.967 \pm 0.039$ \\
part$\to$lund & $0.524 \pm 0.188$ & $0.542 \pm 0.300$ & $0.679 \pm 0.166$ & $0.799 \pm 0.077$ & $0.918 \pm 0.042$ \\
part-self & $0.460 \pm 0.225$ & $0.582 \pm 0.176$ & $0.587 \pm 0.170$ & $0.790 \pm 0.081$ & $0.900 \pm 0.082$ \\
\end{tabular}
\end{ruledtabular}
\end{table*}

\begin{figure*}[t]
\centering
\includegraphics[width=0.9\textwidth]{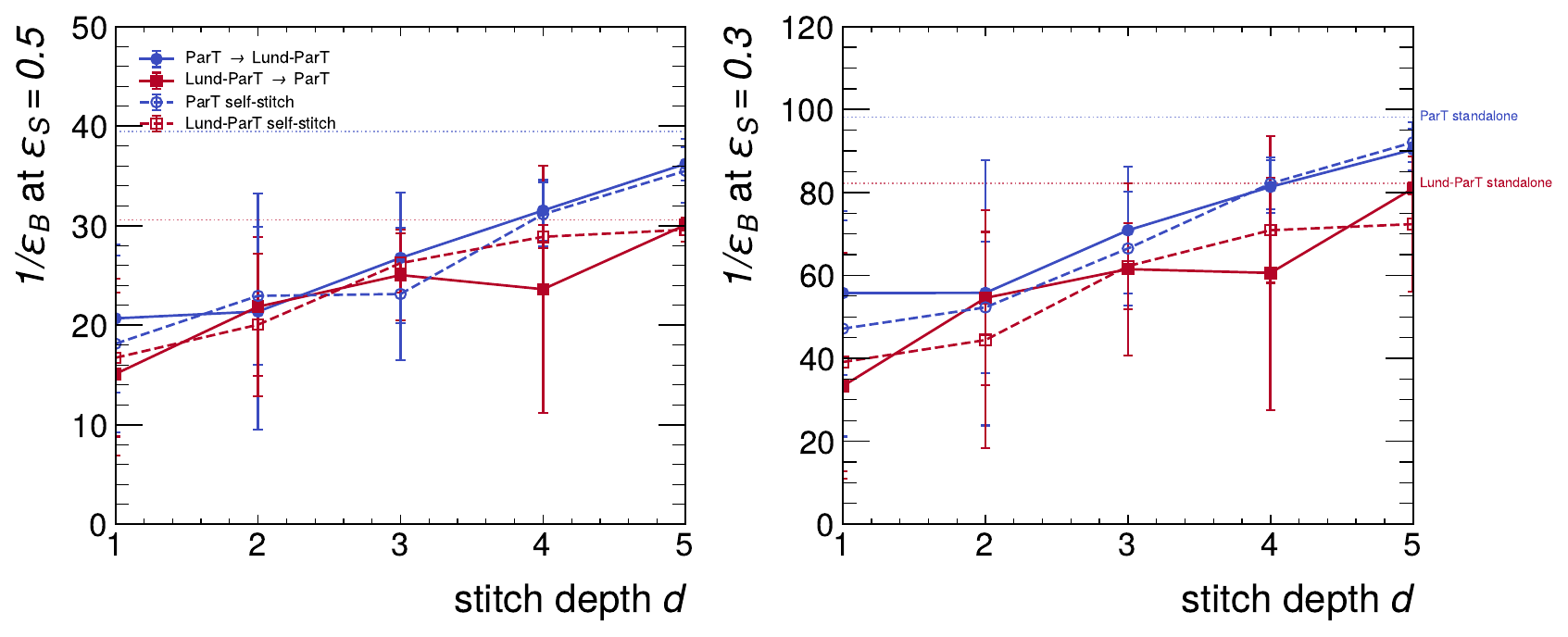}
\caption{Absolute background rejection at $\varepsilon_S = 0.5$ and $0.3$ for all four stitching conditions on quark/gluon tagging, primary-branch run, versus depth. Horizontal lines mark each network's own standalone value. Compare to Fig.~\ref{fig:absolute} for top tagging.}
\label{fig:qg-absolute}
\end{figure*}

\paragraph{Consumer-asymmetry test.} Averaging the self-minus-cross difference in raw rejection over $d=1,\dots,5$ per seed, then over the two directions, gives a mean advantage for ParT's downstream over Lund-ParT's of $-1.1 \pm 2.7$ rejection units ($t=-0.95$, $p=0.395$, $n=5$ seeds). The sign is negative and the result is not significant -- the opposite of the top-tagging result (Sec.~\ref{sec:bothnorm}), where the same test gave $+8\pm7$ ($p=0.062$).

\subsection{Geometric alignment}

Linear CKA and SVCCA between ParT's and Lund-ParT's class tokens, computed on a fixed probe sample independently of any stitching or classification objective. The depth index here is the layer at which a class token is read from a standalone network, so unlike the stitch sweep it remains defined at $d=6$.

\begin{table}[htbp]
\caption{Geometric alignment between ParT's and Lund-ParT's class tokens, quark/gluon tagging.}
\label{tab:qg-geom}
\footnotesize
\begin{ruledtabular}
\begin{tabular}{lcc}
 & CKA & SVCCA \\
\hline
primary branch, $d{=}1$ & $0.639\pm0.118$ & $0.769\pm0.091$ \\
primary branch, $d{=}6$ & $0.884\pm0.008$ & $0.925\pm0.034$ \\
full tree, $d{=}1$ & $0.379\pm0.119$ & $0.670\pm0.111$ \\
full tree, $d{=}6$ & $0.919\pm0.005$ & $0.948\pm0.026$ \\
\end{tabular}
\end{ruledtabular}
\end{table}

\begin{figure}[htbp]
\centering
\includegraphics[width=0.85\linewidth]{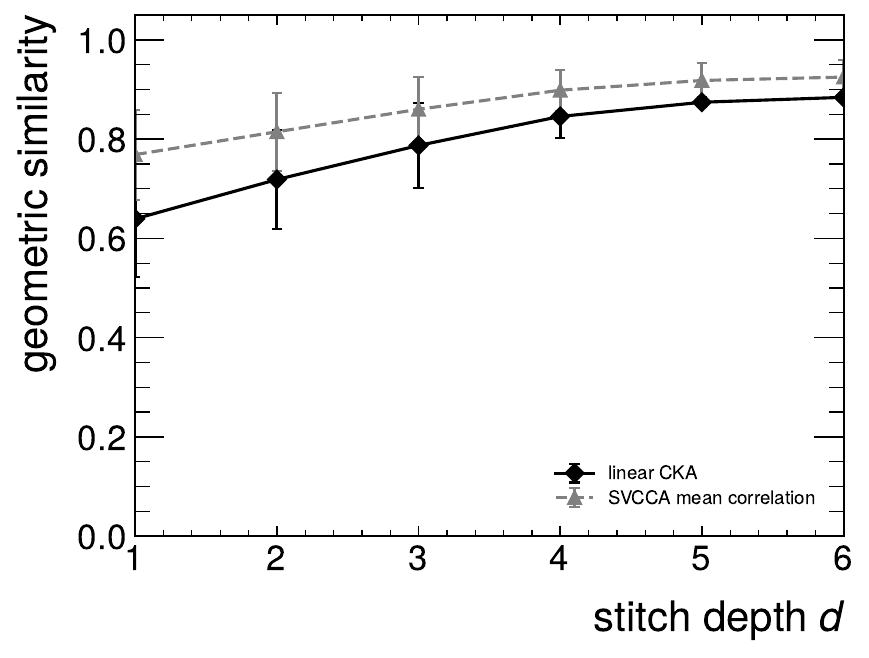}
\caption{Linear CKA and SVCCA mean canonical correlation between ParT's and Lund-ParT's class tokens versus depth, quark/gluon tagging. Compare to Fig.~\ref{fig:cka} for top tagging.}
\label{fig:qg-cka}
\end{figure}

\subsection{Jet-level localization}

Of the $100{,}016$ gluon jets in the test split, standalone ParT rejects $97{,}575$ at its own $\varepsilon_S = 0.5$ threshold. Table~\ref{tab:qg-localization} counts those the stitched model fails to reject that standalone ParT would have, with $\sqrt{N}$ counting uncertainties.

\begin{table}[htbp]
\caption{Leakage relative to standalone ParT's rejected set, quark/gluon tagging.}
\label{tab:qg-localization}
\begin{ruledtabular}
\begin{tabular}{llccc}
representation & $d$ & leaked & \% of rejected set & newly rejected \\
\hline
primary branch & 1 & $2607\pm51$ & $2.7\%$ & 643 \\
primary branch & 5 & $1306\pm36$ & $1.3\%$ & 486 \\
full tree & 1 & $5539\pm74$ & $5.7\%$ & 655 \\
full tree & 5 & $897\pm30$ & $0.9\%$ & 400 \\
\end{tabular}
\end{ruledtabular}
\end{table}

Spearman rank correlation with standalone ParT over all gluon jets is $0.812$ ($d=1$) and $0.836$ ($d=5$) for the primary branch, and $0.661$ and $0.907$ for the full tree.

\begin{figure*}[t]
\centering
\includegraphics[width=0.9\textwidth]{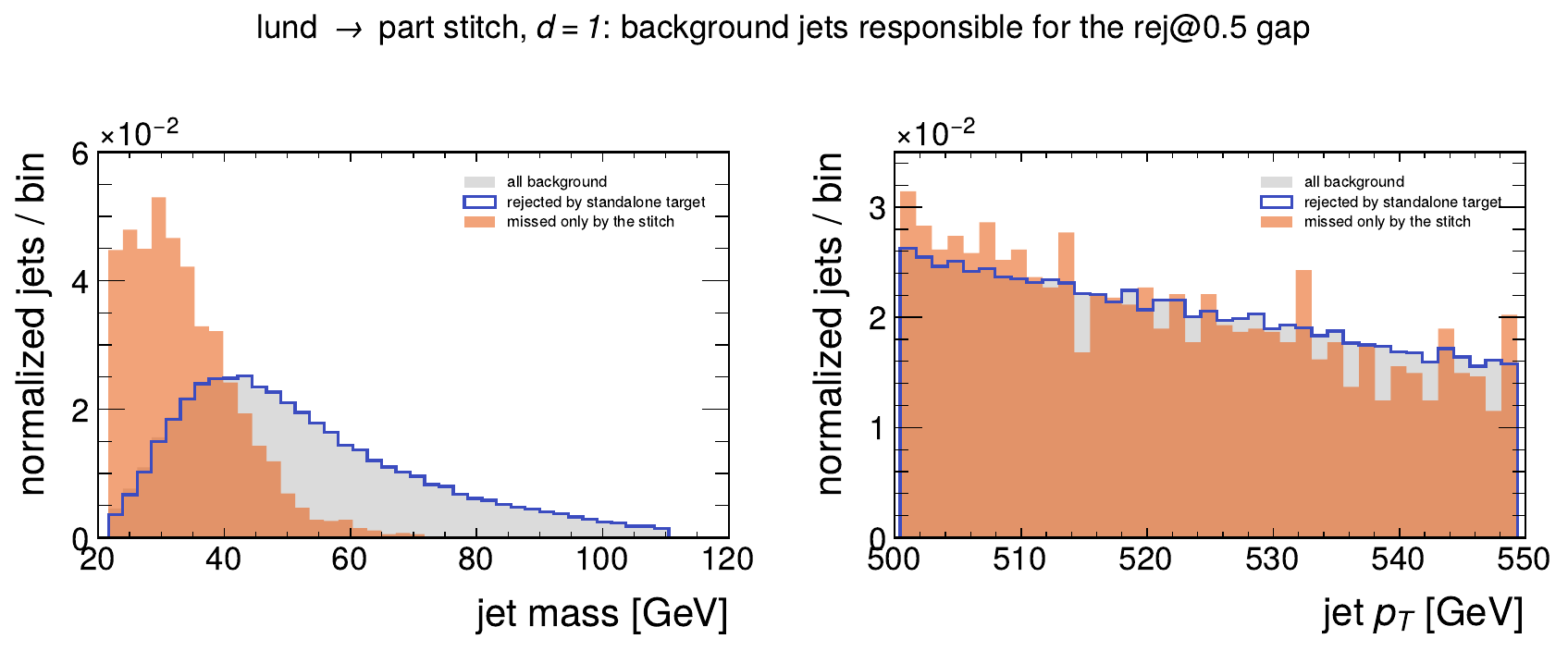}
\caption{Jet mass and $p_T$ for all gluon jets, the subset standalone ParT rejects at its own $\varepsilon_S = 0.5$ threshold, and the subset the lund-to-part stitch at $d=1$ fails to reject that standalone ParT would have.}
\label{fig:qg-leakage-d1}
\end{figure*}

\begin{figure*}[t]
\centering
\includegraphics[width=0.9\textwidth]{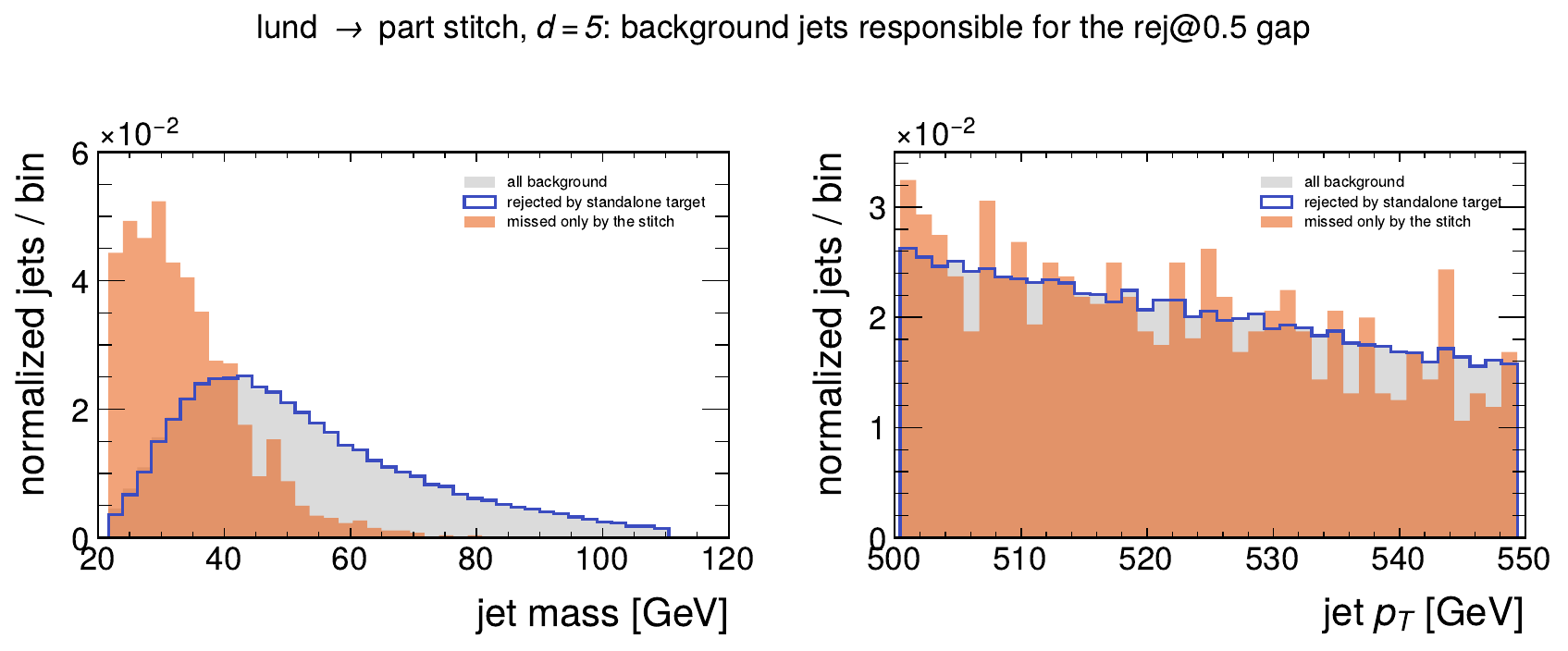}
\caption{As Fig.~\ref{fig:qg-leakage-d1}, at $d=5$.}
\label{fig:qg-leakage-d5}
\end{figure*}

\subsection{Reading the numbers}

Three features of this run should be stated explicitly, because they bear on what can and cannot be concluded from the tables above.

\paragraph{Most of the depth structure is not resolved.} The per-depth stitch measurements carry seed spreads of $20$--$55\%$ in relative terms at $d \le 4$: $15.1\pm8.2$ at $d=1$ and $23.6\pm12.4$ at $d=4$ for the primary branch, for instance. The non-monotonic entries in Tables~\ref{tab:qg-auc} and~\ref{tab:qg-rsrc} -- lund$\to$part dipping at $d=4$, part$\to$lund at $d=2$ -- are consistent with that scatter and should not be read as structure. Only $d=5$, where the spreads fall to a few percent, is measured sharply. This is a substantially noisier regime than the top-tagging study, where per-depth spreads were a few percent throughout.

\paragraph{The consumer asymmetry does not appear here.} In the top-tagging study ParT's downstream extracted at least as much from the Lund representation as Lund-ParT's own did, at every depth. On these jets the ordering is mixed across depths and the seed-level test returns $-1.1\pm2.7$ with $p=0.395$: the opposite sign, and consistent with zero.

\paragraph{The full tree does not dominate the primary branch at every depth.} Standalone, the full tree is ahead, $33.54$ against $30.59$. Through the stitch it is \emph{behind} at $d=1,2,3$ and ahead only at $d=4,5$ (Table~\ref{tab:qg-absolute}), and its leaked population at $d=1$ is larger than the primary branch's, $5539$ against $2607$, shrinking below it only at $d=5$, $897$ against $1306$. The $d=1$ comparison also has the lowest rank correlation of the four, $0.661$. In the top-tagging study the full tree led at every depth and its leaked population was smaller at both depths checked.

\subsection{Physical interpretation}

The comparison that carries the physics is against the same analysis run on the top-tagging reference dataset, at 50\% of that dataset and over the same five seeds, where ParT reaches $290.7\pm13.2$, the primary branch $136.3\pm4.3$, and the full tree $247.7\pm5.3$ at $\varepsilon_S=0.5$ (Table~\ref{tab:standalone}).

\begin{table}[htbp]
\caption{Standalone rejection at $\varepsilon_S = 0.5$ as a fraction of ParT's, on the two datasets.}
\label{tab:qg-crosstask}
\begin{ruledtabular}
\begin{tabular}{lccc}
 & primary/ParT & full tree/ParT & gap closed by tree \\
\hline
quark/gluon & $77.6\%$ & $85.0\%$ & $33\%$ \\
top tagging & $46.9\%$ & $85.2\%$ & $72\%$ \\
\end{tabular}
\end{ruledtabular}
\end{table}

\paragraph{The primary Lund plane is a far better representation here than for top tagging.} On its own it retains $77.6\%$ of ParT's rejection on quark/gluon tagging against $46.9\%$ on top tagging (Table~\ref{tab:qg-crosstask}). This follows the physics of the two tasks. Quark/gluon separation rests on Casimir scaling, $C_A/C_F = 9/4$: gluons radiate more, which shows up as the rate and hardness of emissions along the jet's hardest chain together with the resulting multiplicity. That is precisely what the primary branch samples, since it records $\ln k_T$ and $\ln\Delta R$ at every step of that chain. Top tagging instead requires a localized three-prong structure, two prongs of which sit on branches the primary plane discards. The same omission is close to fatal for one task and only mildly costly for the other.

\paragraph{Restoring the full tree therefore buys much less.} It closes $33\%$ of the gap to ParT here against $72\%$ there. In a quark or gluon jet the secondary branches carry further soft radiation of the same character the primary chain already samples, so they add statistics on a property that was already measured. In a top jet they carry the missing hard prongs. The construction is identical; what it recovers is not.

\paragraph{The residual shortfall against ParT is the same on both tasks.} Even a complete declustering tree sits at $85\%$ of ParT's rejection on either dataset ($85.0\%$ and $85.2\%$). Since the two tasks differ in which structures discriminate, a shortfall this similar is more naturally read as a cost of the Lund encoding itself than of any particular omitted branch: the Cambridge/Aachen sequence is one hierarchical summary of a jet, and what that summary compresses away appears to cost a roughly task-independent fraction of the rejection.

\paragraph{The full tree's advantage requires depth before it appears.} Through the stitch the full tree is behind the primary branch at $d=1,2,3$ and ahead only at $d=4,5$ (Table~\ref{tab:qg-absolute}); its leaked population at $d=1$ is the larger of the two, $5539$ against $2607$, and carries the lowest rank correlation of the four comparisons, $0.661$; and its geometry starts further from ParT's, CKA $0.379$ against $0.639$ at $d=1$, before converging higher by $d=6$, $0.919$ against $0.884$ (Table~\ref{tab:qg-geom}). The roughly 28 additional nodes per jet are therefore not locally readable: they require several rounds of attention before they aggregate into anything discriminating. The information the full tree adds is distributed rather than concentrated.

\paragraph{The consumer asymmetry disappears.} On top tagging, ParT's downstream extracted at least as much from the Lund representation as Lund-ParT's own did, giving $+8.0\pm7.0$ rejection units ($p=0.062$). Here the test returns $-1.1\pm2.7$ ($p=0.395$). The natural reading is that little latent information remains for a stronger downstream to recover once the representation already reaches $77.6\%$ of ParT: the primary branch is near its ceiling on this task, where on top tagging it was far from it.

\begin{figure*}[t]
\centering
\includegraphics[width=0.9\textwidth]{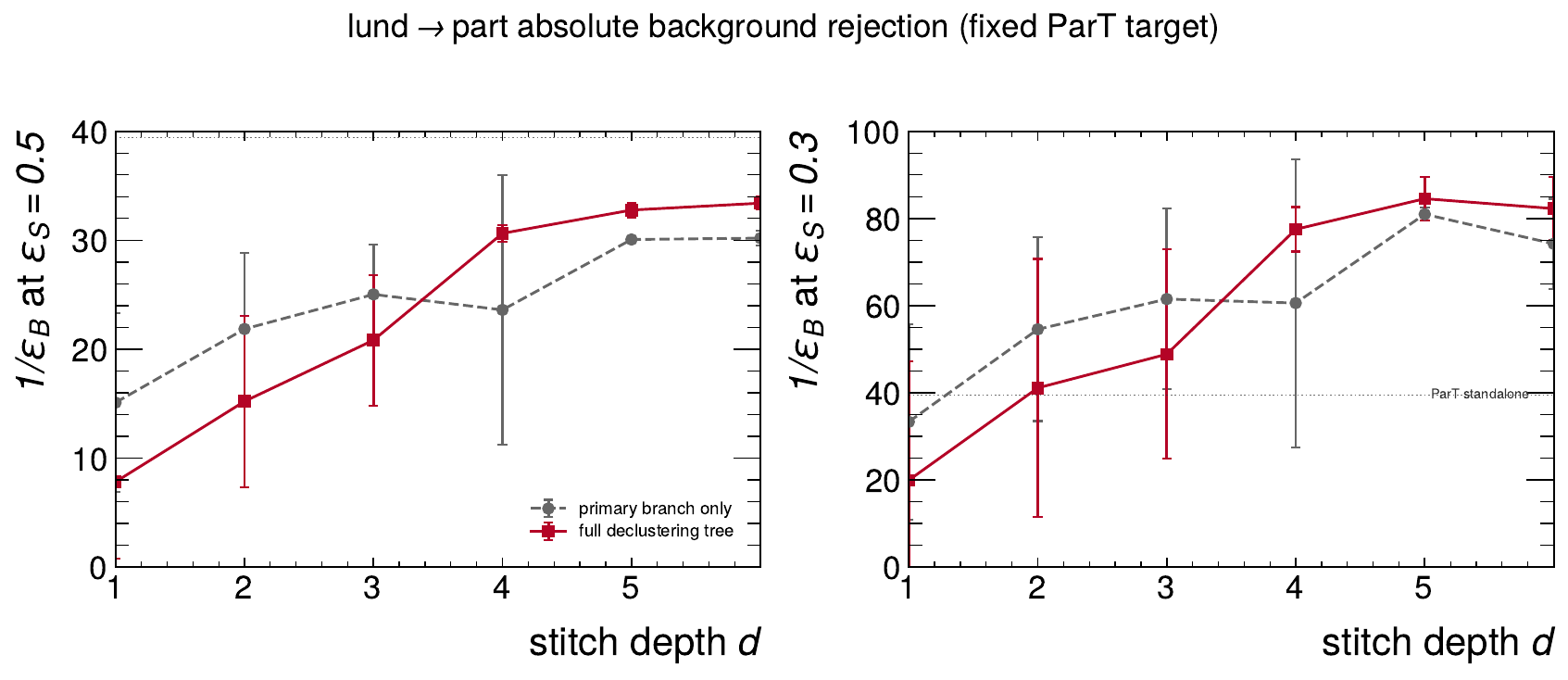}
\caption{Lund-to-part absolute rejection on quark/gluon tagging, primary branch versus full declustering tree, against the same ParT target throughout. The horizontal line marks ParT's own standalone value. Compare to Fig.~\ref{fig:repair} for top tagging.}
\label{fig:qg-repair}
\end{figure*}

\begin{figure*}[t]
\centering
\includegraphics[width=0.9\textwidth]{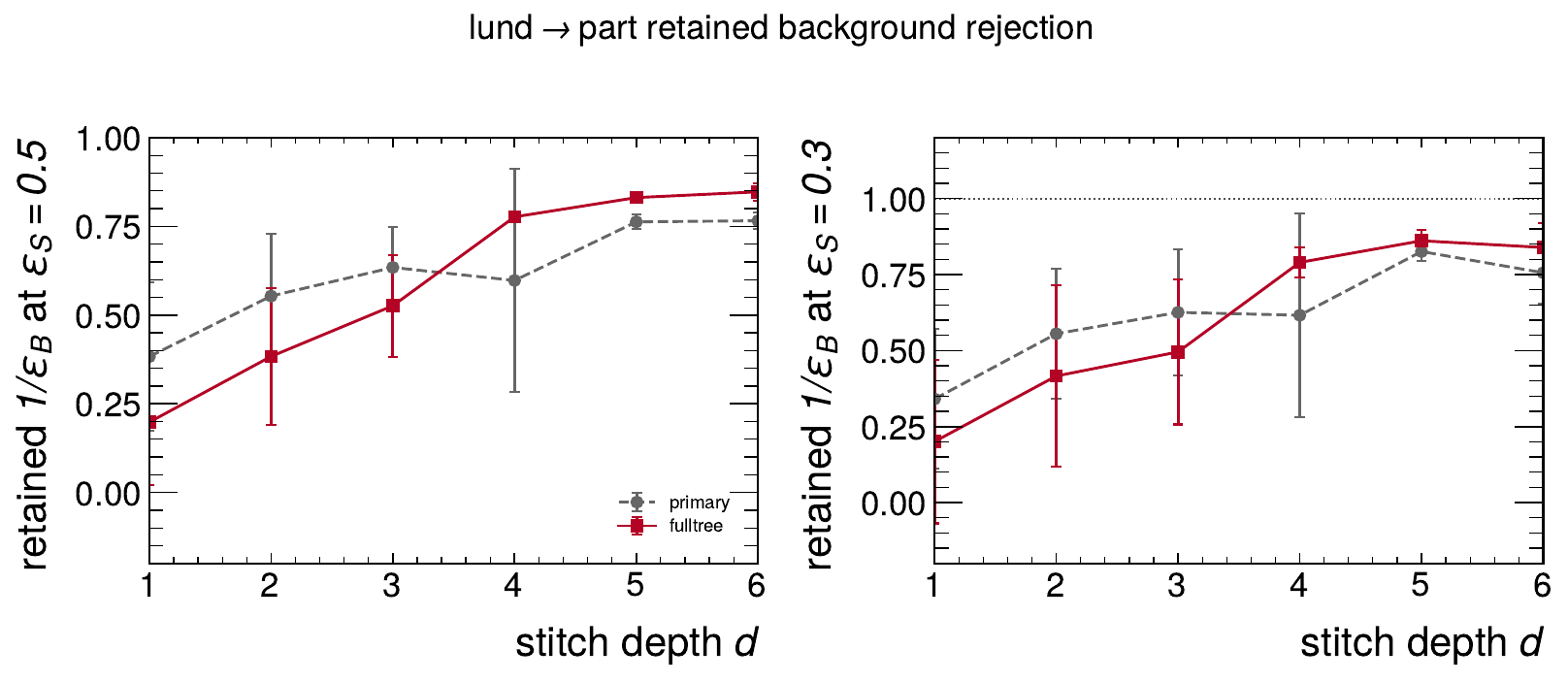}
\caption{Primary branch against full declustering tree across the stitch sweep, quark/gluon tagging.}
\label{fig:qg-comparison}
\end{figure*}

\paragraph{What these readings do not rest on.} The depth-by-depth stitch measurements carry $20$--$55\%$ seed spreads below $d=5$, so the ordering at any individual shallow depth is not resolved on its own; the statement about depth above rests on the direction being consistent across $d=1,2,3$ together with the independent leakage and CKA evidence, not on any single entry. The asymmetry test is underpowered at five seeds, so the absence of a detected asymmetry is not evidence of its absence. Both conclusions compare two datasets at different fractions of their respective reference samples, with the architecture and every hyperparameter held fixed but the jet kinematics differing in $p_T$ reach and multiplicity.

\clearpage
\bibliographystyle{apsrev4-2}
\bibliography{references}

\end{document}